\documentclass[12pt]{article} 
\usepackage[utf8]{inputenc}
\usepackage[english]{babel}

\usepackage{makecell}
\usepackage{textcomp}
\usepackage{amsmath,etoolbox}
\usepackage{amsthm}
\usepackage{amssymb}
\usepackage[left=2.5cm,right=2.5cm,top=3cm,bottom=3cm,bindingoffset=0cm]{geometry}
\usepackage{dsfont}
\usepackage{bm}
\usepackage{leftidx}
\usepackage[makeroom]{cancel}
\usepackage{indentfirst}
\usepackage{underoverlap}
\usepackage{caption}
\usepackage{hyperref}
\usepackage{tikz}
\usepackage{upgreek}
\usepackage{subfig}
\usepackage{subdepth}
\usepackage{graphicx}
\usepackage{tikz-cd}
\usepackage[titletoc,title]{appendix}

\usepackage{adjustbox}
\usepackage{mathtools}
\usepackage{multirow}
\usepackage{xargs}
\usepackage{xstring}

\usetikzlibrary{matrix,arrows,decorations.pathmorphing}
\newenvironment{sloppypar*}{\sloppy\ignorespaces}{\par}
\allowdisplaybreaks 
\newcommand{\lowe}{\text{lowest eigenvalue}}

\newcommand {\sket} [1] {| #1 \rangle}
\newcommand {\sbraket} [2] {\langle #1 | #2 \rangle}
\newcommand {\saxe} [2] {| #1 \rangle\langle #2 |}
\newcommand {\sand} [3] {\langle #1 | #2 | #3 \rangle}

\renewcommand\Re{\operatorname{Re}}
\renewcommand\Im{\operatorname{Im}}

\newcommand{\sdagger}{{\hspace{.9pt}\dagger}}
\DeclareMathOperator{\Tr}{Tr\s{}  }

\newcommand{\psib}{\overline{\psi}}
\newcommand{\bma} {\begin{pmatrix}}
\newcommand{\ema} {\end{pmatrix}}

\newcommand{\iu}{{i\mkern1mu}}

\renewcommand{\d}[1]{\ensuremath{\operatorname{d}\!{#1}}}
\newcommand{\me}{\mathrm{e}}
\newcommand{\s}{\hspace{.08em}}
\newcommand{\hilbdim}{\mathcal{D}}
\newcommandx*\hilbdimK[1][1=]{\mathcal{D}_{K\IfEqCase{#1}{{}{}}[=#1]}}
\newcommandx*\hilbdimKQ[2][1=,2=]{\mathcal{D}_{K\IfEqCase{#1}{{}{}}[=#1],Q\IfEqCase{#2}{{}{}}[=#2]}}

\newcommand{\qubitn}{\mathsf{Q}}

\newcommand{\spa}{\eta}

\newcommand{\spam}{\spa_\text{lower}}
\newcommand{\spaM}{\spa_\text{upper}}
\newcommand{\bsm}{P^+} 
\newcommand{\Modep}{{\mathsf{p}}}
\newcommand{\Moden}{{\mathsf{n}}}
\newcommand{\Modem}{{\mathsf{m}}}
\newcommand{\Modek}{{\mathsf{k}}}
\newcommand{\ccolor}{{\mathsf{c}}}
\newcommand{\acolor}{{\mathsf{a}}}
\newcommand{\scolor}{{\mathsf{s}}}
\newcommand{\rcolor}{{\mathsf{r}}}

\newcommand{\bcolor}{{\mathsf{b}}}
\newcommand{\Model}{{\mathsf{l}}}

\newcommand{\mass}{{\mathrm{m}}}
\newcommand{\occ}{{w}}
\newcommand{\maxoc}{\widetilde{r}} 

\newcommand{\OPa}{a^{\vphantom{\dagger}}}
\newcommand{\OPad}{a^{\dagger}}
\newcommand{\OPb}{b^{\vphantom{\dagger}}}
\newcommand{\OPbd}{b^{\dagger}}
\newcommand{\OPc}{c^{\vphantom{\dagger}}}
\newcommand{\OPcd}{c^{\dagger}}
\newcommand{\OPd}{d^{\vphantom{\sdagger}}}
\newcommand{\OPdd}{d^{\sdagger}}
\newcommand{\numberop}{{\mathcal{N}}}
\newcommand{\formfM}{{\mathcal{F}}}
\newcommand{\FF}{{F}}

\newcommand{\snorm}[1]{\lVert#1\rVert}

\newcommand{\snormmax}[1]{{{\snorm{#1}}_{\text{max}}}}

\newcommand{\Aorb}{\overline{\mathcal{F}}}
\newcommand{\Borb}{\vphantom{\Aorb}\widetilde{\mathcal{B}}}
\newcommand{\Forb}{\vphantom{\Borb}\mathcal{F}}

\newcommand{\collective}{{\xi}}
\newcommand{\collectivestruck}{{\collective_{\text{st}}}}
\newcommand{\Modecollective}{{\upxi}}

\newcommand{\Modeflavor}{{\mathsf{f}}}

\newcommand{\bitprec}{{\mathrm{p}}}

\newcommand{\helicity}{{\lambda}}

\newcommandx*\bound[3][,2=,3=]{#1_{K
\IfEqCase{#2}{{}{}}[,\,#2]}^{
\IfEqCase{#3}{{}{}}[(#3)]
}}

\newcommand{\Qtransf}{{\mathcal{Q}}}

\DeclareMathAlphabet{\mathdutchcal}{U}{dutchcal}{m}{n}
\newcommand{\charge}{{\mathdutchcal{q}}}

\def\ket#1{\left\vert #1 \right\rangle}
\def\bra#1{\left\langle #1 \right\vert}

\newcommand{\be}{\begin{equation}}
\newcommand{\ee}{\end{equation}}
\newcommand{\bp}{\begin{pmatrix}}
\newcommand{\ep}{\end{pmatrix}}
\newcommand{\ben}{\begin{enumerate}}
\newcommand{\een}{\end{enumerate}}

\let\oldFootnote\footnote
\newcommand\nextToken\relax
\renewcommand\footnote[1]{\oldFootnote{#1}\futurelet\nextToken\isFootnote}
\newcommand\isFootnote{\ifx\footnote\nextToken\textsuperscript{,}\fi}

\usepackage[shortcuts]{extdash}

\usepackage{authblk}

\newcommand*\standardbin{+}

\makeatletter

\patchcmd{\env@cases}
  {\quad}
  {{ \ }}
  {}{}

\newcommand*\tabularbin[1]{%
  \mathbin{\mathpalette{\@tabularsym\standardbin}{#1}}%
}

\newcommand*\@tabularsym[3]{%
  \setbox\z@\hbox{$#2#1\m@th$}%
  \hbox to\wd\z@{\hss$#2#3\m@th$\hss}%
}

\newcommand*{\myrulefill}[3][]{%
  \makebox[#2]{#1%
    \leaders\hrule height \dimexpr.5ex+.2pt\relax depth \dimexpr -.5ex+.2pt\relax \hfill
    \enskip{#3}\enskip
    \leaders\hrule height \dimexpr.5ex+.2pt\relax depth \dimexpr -.5ex+.2pt\relax \hfill\kern0pt}
}

\renewcommand\AB@authnote[1]{\rlap{\textsuperscript{\normalfont#1}}}

\makeatother
\author[1]{Michael Kreshchuk}
\author[1]{William M. Kirby}
\author[1]{Gary Goldstein}
\author[1]{Hugo Beauchemin}
\author[1,2]{Peter J. Love\thanks{peter.love@tufts.edu}}
\affil[1]{{\small
~Department of Physics and Astronomy, Tufts University, Medford, MA, 02155, USA
}}
\affil[2]{{\small
~Computational Science Initiative, Brookhaven National Laboratory, Upton, NY 11973, USA
}}
\title{{\Large{\textbf{
Quantum Simulation of Quantum Field Theory\\in the Light-Front Formulation
}}}}
\date{\small \today}

\begin{document}
\pagenumbering{arabic}

\maketitle
\begin{abstract}

Quantum chromodynamics (QCD) describes the structure of hadrons such as the proton at a fundamental level.
The precision of calculations in QCD limits the precision of the values of many physical parameters extracted from collider data. For example, uncertainty in the parton distribution function (PDF) is the dominant source of error in the $W$ mass measurement at the LHC. Improving the precision of such measurements is essential in the search for new physics.

Quantum simulation offers an efficient way of studying quantum field theories (QFTs) such as QCD non-perturbatively. Previous quantum algorithms for simulating QFTs have qubit requirements that are well beyond the most ambitious experimental proposals for large-scale quantum computers. Can the qubit requirements for such algorithms be brought into range of quantum computation with several thousand logical qubits?
We show how this can be achieved by using the light-front formulation of quantum field theory.
This work was inspired by the similarity of the light-front formulation to quantum chemistry, first noted by Kenneth Wilson.
\end{abstract}


\newpage
\tableofcontents
\newpage


\section*{Introduction\label{intro}}
\addcontentsline{toc}{section}{\nameref{intro}}

Feynman first proposed using one quantum system to simulate another~\cite{feynman1982}. A decade later the first general quantum algorithms appeared~\cite{lloyd1996universal,zalka1998simulating,wiesner1996simulations,boghosian1997quantum,meyer1996quantum}, with applications to quantum chemistry~\cite{aspuru2005simulated}, condensed matter~\cite{wu2002polynomial}, and high energy physics~\cite{preskill1}. Quantum simulation is now recognized as a significant future application of quantum computation~\cite{NSFWSR,BESReport,ASCRReport}, especially in the context of near-term devices. Quantum algorithms for quantum simulation with almost optimal scaling are now known~\cite{berry2017exponential,low2017optimal,low2016hamiltonian,berry2015hamiltonian,berry2015simulating}. Applications of these methods to condensed matter and quantum chemistry are well-developed theoretically~\cite{babbush2018encoding,kivlichan2017bounding,babbush2017low,babbush2016exponentially,babbush2015chemical,BK2015,Toloui,32JCP,yung2014transistor,kassal2008polynomial}, and experiments have been performed on many different quantum architectures~\cite{lanyon2010towards,li2011solving,wang2015quantum,o2016scalable,shen2017quantum,2016sangatti,paesani2017,kandala2017hardware,Hempel2018,martinez2016real,garcia2015fermion,marcos2014two,peruzzo2014variational,cloudnucleus,colless18a,nam19a,kokail2018self,kandala19a}.

Quantum simulation of relativistic quantum field theory poses new challenges. Among these challenges are the absence of any fixed particle-number formulation of relativistic quantum theory, multiple particle types with varying statistics, complicated interactions and observables, nontrivial vacuum structure and gauge invariance. Nevertheless, quantum simulation is the only efficient approach to the study of general QFTs in the non-perturbative regime.

Quantum simulation of high energy physics can be approached via analog simulation of lattice gauge theories in cold atoms or ions~\cite{wiese2014towards,zohar2015quantum,zohar2013cold,zohar2013quantum,gonzalez2017quantum,zhang2016fermion,opportunities}, analog simulation using continuous variable quantum systems~\cite{barrett2013simulating, marshall2015quantum}, or digital simulation of quantum field theories using conventional qubits and gates~\cite{preskill1,2017arXiv171104006H,preskill2,bauer2019}.  Theoretical proposals for digital quantum simulation of quantum field theory~\cite{preskill1,preskill2,2011arXiv1112.4833J,Jordan2018bqpcompletenessof,discretizing2020} were followed by experimental implementations of simple models such as the Schwinger model in \mbox{$1+1D$~\cite{martinez2016real,muschik2017u}.}

However, \emph{ab initio} digital quantum simulation of general QFTs using existing techniques requires hundreds of thousands of logical qubits~\cite{Lamm:2019uyc,preskill2018simulating}. This is far beyond what is required for Shor's algorithm, which has been the subject of serious architectural studies estimating requirements of several thousand logical qubits~\cite{gidney2019factor,whitney2009fault}.

Can the ideas explored for quantum simulation of quantum chemistry be used to enable simulation of QFT on quantum computers with several thousand logical qubits? Fortunately we can be guided by Wilson, who suggested~\cite{wilson1990ab} that the light-front formulation of QFT~\cite{RevModPhys.21.392,infiniteweinberg,CARBONELL1998215,Susskind:1967rg} is amenable to the orbital representations used in chemistry. The light-front formulation is now well-developed~\cite{HILLER201675}. Among its notable advantages are a trivial vacuum and the absence of ghost fields. The linearity of equations of motion further reduces the number of independent variables. While the discretized light-front quantization (DLCQ)~\cite{BRODSKY1998299} provides a natural framework for simulating fundamental interactions \emph{ab initio}~\cite{varybasis}, the basis light-front quantization (BLFQ)~\cite{varybasis} approach is well suited for constructing effective theories; in the present work, we focus on the former.

The main goal of the current paper is to demonstrate that the light-front formulation is advantageous for digital quantum simulation. First, the second-quantized form of the light-front Hamiltonian permits a highly efficient encoding scheme, with qubit requirements scaling logarithmically with the spacetime dimension. This reduces qubit requirements by several orders of magnitude: for example, the qubit numbers for the calculation in~\cite{Lamm:2019uyc} are reduced from $\sim\!400000$ to $\sim\!1360$ qubits. Second, the Hamiltonian in this encoding is sparse\footnote{Here, \emph{sparsity} refers to the maximum number of off-diagonal elements in any column (or row, since the Hamiltonian is Hermitian).}~\cite{VaryAbInitio,BRODSKY1998299}, so one can employ sparsity-based simulation algorithms that are almost optimal in all parameters~\cite{berry2017exponential,berry2015hamiltonian}.
Third, the light-front approach is well-adapted to calculation of static quantities such as parton distribution functions, hadronic tensors, form factors, and decay constants~\cite{Collins:2011zzd}. All of these can be calculated directly from the light-front wave function, within the Fock space sector with some fixed light-front momentum\footnote{this corresponds to switching to the Drell-Yan-West frame~\cite{BRODSKY1998299,frame,varybasis,Gutsche2017,basislightmesons,partondist}.}.
This leads to a simple form for the corresponding qubit measurement operators.

These advantages apply to \emph{any} light-cone formulation of a relativistic field theory. In the present work we focus on the DLCQ approach, which amounts to solving the fundamental theory in the plane-wave basis~\cite{BRODSKY1998299}. Within a complementary approach, BLFQ, one instead chooses the set of basis functions and effective interactions that are most suitable for a particular problem of interest. Quantum simulation algorithms based on this latter technique will be investigated in subsequent work.


We focus on the quantum computation of static quantities, which in our context refers to single-particle properties such as parton distribution functions (PDFs)~\cite{Gutsche2017,BRODSKY1998299,partondist}, form factors~\cite{frame}, and decay constants~\cite{basislightmesons}. PDFs constitute the dominant source of uncertainty on multiple cross section predictions at the LHC~\cite{lin2018parton,detmold2019hadrons}, substantially affecting the reach of searches for new physics at high final state masses. They affect experimental results as they limit the accuracy to which precision electroweak observables can be extracted from LHC data. To give an example, the difference in the $W^+$ and $W^-$ masses as measured at the LHC is $29.2\pm 28$~MeV, with the PDF uncertainty accounting for $23.9$~MeV, or $85\%$, of the uncertainty~\cite{aaboud2018measurement}.

Lattice calculation techniques on a classical computer have been very successful at calculating static properties of quantum chromodynamics (QCD) in non-perturbative regimes. For example, they provide the most precise way of determining heavy quark masses, improving the uncertainty on \mbox{c-quark} mass over non-lattice predictions by more than a factor of two~\cite{highpreccb,precisecb}. Similar improvement has been obtained on the estimate of the strong coupling constant~$\alpha_S$~\cite{highpreccb}. This has a direct impact on the high-energy collider physics programs, as the parametric uncertainty on heavy-quark masses is the dominant uncertainty on the determination of the branching ratio of most of the Higgs boson decay channels. However, the static property that has the largest impact on physics at the current energy frontier is PDF.

Exploiting the factorization of short distance physics from universal large distance phenomena, the PDFs used at the LHC are obtained from a parameterized asymptotic form at low resolution~($\Qtransf^2$), perturbatively evolved to higher resolution at which cross sections in the partonic center-of-mass system are calculated. This \mbox{low-$\Qtransf^2$} parametric form is largely responsible for the uncertainty in the knowledge of the PDF. A more precise prediction of the PDF in such a regime, and eventually at higher~$\Qtransf^2$, would therefore significantly improve the precision of theoretical predictions and of many experimental measurement results at LHC energies.

Currently, the dominant approach for performing \emph{ab initio} QCD calculations in the strong coupling regime is lattice QCD (LQCD)~\cite{lin2018parton,detmold2019hadrons}.
Within the traditional approach to LQCD, one evaluates the PDFs indirectly, by calculating the matrix elements of local twist-two operators~\cite{lin2018parton,detmold2019hadrons}. From a sufficient number of these operators the Mellin moments of PDFs can be reconstructed. In practice, one is limited to the first three moments because power-divergent mixing between the operators occurs due to the reduced symmetries of the spacetime lattice.
Considerable progress has been made recently by applying large momentum effective theory techniques~\cite{Ji_lattice}. \emph{Quasi-distributions}~\cite{Constantinou,quasi1,quasi2,quasi3,quasi4,quasi5,quasi6,quasi7,quasi8,quasi9} allow for the equivalent of higher moment calculations, by matching higher moment calculations to the effective field theory.
More recent approaches include finding PDFs from the hadronic tensor~\cite{hadronic1,hadronic2,hadronic3,hadronic4,hadronic5} and Compton amplitude~\cite{compton1,compton2,compton3,compton4,compton5}, using \emph{pseudo}-PDFs~\cite{pseudo1,pseudo2,pseudo3,pseudo4,pseudo5,pseudo6,pseudo7}, and calculating \emph{good lattice cross-sections}~\cite{good1,good2,good3}. As noted above, these calculations are at present not sufficient to reduce the theoretical uncertainty due to the PDF in many high energy physics measurements such as~\cite{aaboud2018measurement}.


Lattice QCD is based on path integral quantization, and thus requires Wilson gluon lines and loops to maintain the color gauge invariance. Finite size lattices with periodic boundary conditions constrain the simulations (as do the prescriptions for fermion sources), causing the fermion doubling problem. The numerical sign problem severely complicates Monte-Carlo sampling in strongly interacting fermionic systems. On the other hand, the second-quantized approach in light-front formulation avoids Wilson loops and gauge group discretization, but eventually must treat the gluon fields on an equal footing with the quark fields and their interactions.

Previous work on digital simulation of QFT has mainly focused on dynamic quantities like scattering cross sections~\cite{preskill1,preskill2,2011arXiv1112.4833J,Jordan2018bqpcompletenessof}. The possibility of studying parton physics on quantum computers was first explored in~\cite{Lamm:2019uyc}. These authors proposed an algorithm for calculating PDFs and the hadronic tensor of the massive Thirring model, based on equal-time quantization of the lattice Hamiltonian. However, because these approaches are based on an equal time lattice formulation of QFT they lead to daunting qubit requirements. 

We study the computation of static quantities by digital quantum simulation in the light front formulation.
In Sec.~\ref{themodelsection}, we review the light-front treatment of a simple \mbox{${1+1}D$} model containing coupled fermion and scalar fields~\cite{PhysRevD.7.1133,pauli1,pauli2}.
In the front form, the Hamiltonian matrix of the field theory quantized in a box is block-diagonal. Each finite size block approximates the Hilbert space of the theory with a certain precision, the so-called \emph{harmonic resolution}. The harmonic resolution therefore plays the same role as the number operator in for example, simulations of quantum chemistry. For this model, we introduce an analogue of the QCD parton distribution function, and propose an algorithm for its calculation.
In Sec.~\ref{algosec}, we present the algorithm for calculating this quantity on a quantum computer, and provide detailed resource estimates for qubit and gate count.
In Sec.~\ref{qcdlf}, we discuss the generalization to ${3+1}D$ QCD, estimate the required qubit resources, and describe how to calculate PDFs, decay constants, and form factors using our algorithm.



\section{Light-front Quantization of the \texorpdfstring{$1+1D$}{1+1D} Yukawa Model\label{themodelsection}}

We now consider a simple \mbox{${1+1}D$} model with Lagrangian~\cite{pauli1,pauli2}
\begin{equation}
    \label{Lagrangian}
    \mathcal{L} = \dfrac{1}{2} (\partial \phi)^2 - \dfrac{1}{2} \mass_B^2\phi^2 + \iu \psib \gamma^\mu \partial_\mu \psi - \mass_F \psib \psi - \lambda \phi \psib \psi \ .
\end{equation}
Here $\phi$ and $\psi$ are mutually interacting real boson and fermion fields. As in QCD, due to \emph{confinement} emerging at low energies, the eigenstates of the interacting theory can be thought of as composite particles~--- \emph{bound states} which are made of quanta of fields $\phi$ and $\psi$, the \emph{partons}. We will introduce analogues of the QCD parton distribution functions in this model.

In~\cite{preskill1,preskill2,Lamm:2019uyc}, authors studied algorithms based on equal-time quantization and spatial discretization of the wave function. We instead use light-cone coordinates $x^\pm$ and work with the second-quantized formulation of the theory known as the discretized light-cone quantization (DLCQ)~\cite{BRODSKY1998299}.

\subsection{The model \label{lfc}}

\begin{sloppypar}
Equal-time coordinates describe the Minkowskian spacetime as seen by a massive observer. The $1+1D$ metric and gamma matrices can be chosen as ${g_{00}=-g_{11}=1}$, ${g_{01}=g_{10}=0}$, ${\gamma^0 = \sigma_3}$, ${\gamma^1 = \iu \sigma_2}$ where $\sigma_3$ and $\sigma_2$ are the Pauli matrices in the standard basis.
Light-front~(LF) coordinates are obtained by performing the following coordinate transformation:
\end{sloppypar}
\begin{equation}
    \label{coords}
    x^\pm = x^0 \pm x^1
    \ ,
\end{equation}
\begin{sloppypar*}
thus switching to the so-called light-front time ($x^+$) and distance ($x^-$). Physically, we may think of this as describing the experience of a massless observer. The metric is ${g^{++}=g^{--}=0}$, ${g^{+-}=g^{-+}=2}$, and the gamma matrices are defined as ${\gamma^\pm=\gamma^0\pm\gamma^1}$. The only independent variables in the LF formulation of the theory~\eqref{Lagrangian} are the fields $\phi$ and $\psi^{(+)}$, where ${\psi^{(\pm)}=\Lambda^{(\pm)}\psi=\frac{1}{4}\gamma^{\pm}\gamma^{\mp}\psi}$. Here $\Lambda^{(\pm)}$ act on the spinor field as projectors, since ${(\Lambda^{(\pm)})^2=\Lambda^{(\pm)}}$ and ${{\Lambda^{(+)}+\Lambda^{(-)}=1}}$~\cite{pauli1}.
\end{sloppypar*}

For a system quantized in a box $x^-\in(-L,L)$, the plane wave expansions of the free fields are
\begin{subequations}
\label{freef}
\begin{alignat}{99}
    \label{philf}
    \phi(x^+,\,x^-) &= \sum\limits_{\Moden=1}^\Lambda \dfrac{1}{\sqrt{4\pi \Moden}}
    &&\Bigl( \OPa_\Moden \me^{-\iu \Modep^\mu_\Moden x_\mu} &&+ \OPad_\Moden \me^{\iu \Modep^\mu_\Moden x_\mu} &&\Bigr) \ &&,\\
    \label{psilf}
    \psi^{(+)}(x^+,\,x^-) &= \dfrac{u}{\sqrt{2L}} \sum\limits_{\Moden=1}^\Lambda
    &&\Bigl( \OPb_\Moden \me^{-\iu \Modep^\mu_\Moden x_\mu} &&+ \OPdd_\Moden \me^{\iu \Modep^\mu_\Moden x_\mu} &&\Bigr) \ &&,
\end{alignat}
\end{subequations}
where $\Lambda$ is the momentum cutoff and $u$ is a \emph{momentum-independent} spinor normalized to unity (unlike in equal-time quantization, where $u_\Moden$ depends on the momentum quantum number $\Moden$). Following~\cite{pauli1,pauli2}, in~\eqref{freef} we impose periodic boundary conditions.
The discretized momenta and energies of the free particles are
\begin{equation}
    \label{discmomenerg}
    \Modep^+_\Moden = \dfrac{2\pi}{L} \Moden
    \ ,\qquad
    \Modep^-_{\Moden} = \dfrac{\mass^2}{\Modep^+_\Moden}
    \ ,\qquad
    \Moden=1,2,3\ldots\,,\,\Lambda\quad,
\end{equation}
where $\mass$ is either the boson or fermion bare mass.
The creation and annihilation operators obey canonical commutation relations: ${[\OPa_\Modem, \OPad_\Moden] = \delta_{\Modem\Moden}}$, ${\{\OPb_\Modem, \OPbd_\Moden\} = \delta_{\Modem\Moden}}$, ${\{\OPd_\Modem, \OPdd_\Moden\} = \delta_{\Modem\Moden}}$.

When quantizing in equal-time ${(x^0,\,x^1)}$ coordinates, a complete set of commuting observables (CSCO) for the theory is given by the charge $Q$, momentum $P$, and energy $E$. Under the transformation~\eqref{coords} to LF coordinates, these become~${P^\pm = E \pm P}$. The charge $Q$, $P^+$, and $P^-$ form a CSCO in the light-cone coordinates~\cite{changrootyan}.

The dimensionless operators $K$ (the so-called \emph{harmonic resolution}) and $H$ (which we shall call the Hamiltonian) are defined by ${P^+=\frac{2\pi}{L}K}$, ${P^-= \frac{L}{2\pi}H}$.
In terms of these light-front operators, the invariant mass operator of the theory can be expressed as
\begin{equation}
    \label{MM}
    M^2
    = E^2 - P^2
    = P^+ P^- = KH \ .
\end{equation}
A study of bound state masses and their renormalization was performed in~\cite{pauli2}. We defer this discussion until Sec.~\ref{massrenorm}.

Note that as one switches from $P^+$ and $P^-$ to the dimensionless operators $K$ and $H$,
the particular value of $L$ may only become important at the stage of converting from light-cone coordinates to equal-time quantities. As it follows from eq.~\eqref{MM}, the value of $L$ is irrelevant for calculating the mass spectrum. As we shall see later, neither will it enter the expression for parton distribution functions (which is to be expected, since the latter describe the \emph{relative} parton momentum distributions within the bound state).

The second-quantized expressions for $H$, $K$, and $Q$ in terms of the creation and annihilation operators are obtained from Lagrangian~\eqref{Lagrangian} by means of the Noether procedure~\cite{pauli1}. The charge and harmonic resolution are:
\begin{equation}
    Q = \sum \limits_\Moden (\OPbd_\Moden \OPb_\Moden -\OPdd_\Moden \OPd_\Moden)
    \ ,\qquad
    K = \sum \limits_\Moden \Moden (\OPad_\Moden \OPa_\Moden + \OPbd_\Moden \OPb_\Moden +\OPdd_\Moden \OPd_\Moden) \ .
\end{equation}

The Hamiltonian $H$ is a sum of four types of terms:
\begin{equation}
    \label{ham}
    H = H_M + H_V + H_S + H_F
    \ .
\end{equation}
$H_M$ is a (diagonal) mass term, while $H_V$, $H_S$ and $H_F$ contain a number of interaction terms qubic and quartic in creation and annihilation operators (see App.~\ref{appham}).

The elements of the Fock space are labeled by orbital occupancies for the fermionic, antifermionic, and bosonic degrees of freedom:
\begin{equation}
    \label{fketlf}
    \begin{gathered}
    \sket{\{\widehat{n}_j,\widehat{\occ}_j\}}
    = \sket{
    {n_1}^{\occ_1},{n_2}^{\occ_2},\ldots,{n_N}^{\occ_N};
    {\overline{n}_1}^{\overline{\occ}_1},{\overline{n}_2}^{\overline{\occ}_2},\ldots,{\overline{n}_{\overline{N}}^{\overline{\occ}_{\overline{N}}}};
    \widetilde{n}_1^{\widetilde{\occ}_1},\widetilde{n}_2^{\widetilde{\occ}_2},\ldots,\widetilde{n}_{\widetilde{N}}^{\widetilde{\occ}_{\widetilde{N}}}
    }
    \ ,\\
    n_j,\overline{n}_j,\widetilde{n}_j  = 1,\,2,\ldots,\Lambda
    \quad,\qquad
    \occ_j,\overline{\occ}_j\in\{0,1\}
    \ ,\qquad
    0\leq\widetilde{\occ}_j\leq
    \lfloor
    \Lambda/\widetilde{n}_j
    \rfloor
    \ .
\end{gathered}
\end{equation}
In eq.~\eqref{fketlf} we only list modes with non-zero occupancies, the hat is used to collectively denote all the particle species.

The crucial fact is that the spectrum of the operator $P^+$ is bounded from below, unlike that of the equal-time momentum $P$. In the equal time formulation, in an inertial reference frame, the Fock space sector of any fixed total momentum contains an infinite number of multi-particle states with that momentum, since those states can contain arbitrary numbers of left- and right-moving particles whose momenta cancel each other. Therefore, in order to obtain a Hilbert space of a finite dimension, one has not only to introduce a momentum cutoff, but also to limit the number of bosonic quanta in a Fock state.

To see how this changes in the light-front, consider an observer moving at the speed of light to the left in equal-time coordinates. In the light-front formulation, this observer has constant light-front coordinate (i.e., is stationary), so to the observer all massive particles appear to be moving to the right, and have positive light-front momentum. Therefore, there can be no cancellation of momenta due to left- and right-moving particles. This implies that in a theory quantized in a box there exists a finite number of states with a given value of~$K$. Thus, by restricting to a particular value of~$K$ one naturally obtains a finite-dimensional Hilbert space without the need to cut off the dimension of the Hilbert space by hand. For a fixed eigenvalue of~$Q$, it turns out that the blocks of the Hamiltonian~$H$ corresponding to larger eigenvalues of~$K$ represent the Hilbert space of the system with a higher resolution.

Within a block of fixed harmonic resolution, the Hamiltonian is proportional to the mass matrix, eq.~\eqref{MM}.
Diagonalization of a \mbox{fixed-$K$} block of~$M^2$ gives a set of bound states $\sket{\bound{\Psi}[s][]}$ with masses $\bound{M}[s][]$~--- these are the physical states of the interacting theory:
\begin{equation}
    \label{msquaredop}
    M^2 \sket{\bound{\Psi}[s][]} = K H \sket{\bound{\Psi}[s][]} = \bigl(\bound{M}[s][]\bigr)^2 \sket{\bound{\Psi}[s][]}
    \ .
\end{equation}
Increasing $K$ results in considering more bound states, with higher resolution. Each state ${s=s^*}$ first appears at some $K_{s^*}$, and is also contained in all the blocks with ${K>K_{s^*}}$. By diagonalizing Hamiltonian blocks of relatively small $K$ one can get a good idea of the general form of the spectrum (see Fig. 1 in \cite{pauli2} and the accompanying discussion).

For a fixed $K$, the lowest eigenvalues of the mass matrix in the $Q=0$ and $Q=1$ sectors correspond to the physical (renormalized) boson and fermion masses. This gives a constraint (the so-called \emph{renormalization condition}), obtained by insisting that these physical masses match their known empirical values. From this constraint we determine the bare masses appearing in the Hamiltonian matrix, which produces the rest of the physical spectrum upon diagonalization; see the discussion in Sec.~\ref{massrenorm}. 

Although the momentum cutoff $\Lambda$ in~\eqref{fketlf} is not used to truncate the Hamiltonian matrix it corresponds to, it explicitly appears in the Hamiltonian due to the presence of the so-called \emph{self-induced inertias} (see App.~\ref{appham}). These play an important role in the mass renormalization~\cite{pauli2}, and are related to vacuum polarization and self-energy terms in the equal-time quantization~\cite{infiniteweinberg}.
In the next section we will show how the wave functions of mass eigenstates can be used to calculate the analogues of QCD PDFs in this model.


\subsection{Parton distribution functions\label{pdfs}}

The light-front approach to QCD is appealing because numerous quantities of practical interest, such as PDFs, elastic form factors, and decay constants can be calculated directly from the light-front wave function~\cite{BRODSKY1998299,varybasis}. The PDF, $f_\ell(x)$, represents the probability of finding a parton of type $\ell$ carrying a certain momentum fraction ${x=\Modep_\Moden^+/\bsm=\Moden/K}$,
where $0<x\leq1$,
inside a bound state (hadron) with total light-front momentum $\bsm = 2\pi K/L$. Given a bound state of the interacting theory, the PDF can be calculated as an expectation value of the single mode number operator summed over all the quantum numbers other than the longitudinal momentum~\cite{Collins:2011zzd,Bouchiat:1971mj,Soper:1976jc} (see also Sec.~\ref{pdfsinlf}). However, since in our model the longitudinal momentum is the \emph{only} quantum number, the PDFs of a particular bound state $\sket{\bound{\Psi}[][]}$ can be calculated simply as
\begin{equation}
    \label{pdfour}
    f_\ell (x)
    =f_\ell({\Modep^+_\Moden}/{\bsm})
    =f_\ell({\Moden}/{K})
    = \sand{\bound{\Psi}[][]}{\numberop_\ell}{\bound{\Psi}[][]}
    \ ,
\end{equation}
with the number operators of different parton species given by
\begin{equation}
    \label{numberoperatorsour}
    \numberop_f(\Moden/K) =
    \OPbd_\Moden \OPb_\Moden
    \ ,\qquad
    \numberop_a(\Moden/K) =
    \OPdd_\Moden \OPd_\Moden
    \ ,\qquad
    \numberop_b(\Moden/K) =
    \OPad_\Moden \OPa_\Moden
    \ .
\end{equation}
These define the number of partons carrying momentum fraction $x=\Moden/K$ inside a hadron studied at harmonic resolution~$K$. Measuring the expectation values as in~\eqref{pdfour} results in evaluating PDFs at $K$ points:
${x={1}/{K},{2}/{K},\ldots,1}$.
For a properly normalized state (with ${\sbraket{\bound{\Psi}[][]}{\bound{\Psi}[][]}=1}$) of total charge~$Q$, the normalization of PDFs is given by
\begin{equation}
    \label{pdfnormalizationour}
    \begin{gathered}
    \sum_{\Moden=1}^{K} \Moden \bigl[f_f(\Moden/K) + f_a(\Moden/K) + f_b(\Moden/K)\bigr] = K
    \ ,\\
    \sum_{\Moden=1}^{K} \bigl[\charge_f f_f (\Moden/K) + \charge_a f_a (\Moden/K) \bigr] =
    \charge_f \sum_{\Moden=1}^{K} \bigl[f_f (\Moden/K) - f_a (\Moden/K) \bigr] =
    Q
    \ ,
    \end{gathered}
\end{equation}
which reflects that fact that momenta and charges of partons should add up to those of the hadron.

The PDF is also a function of the probing scale $\Qtransf^2$, which is the magnitude of momentum exchanged in a scattering process. The probing scale $\Qtransf^2$ can be introduced by imposing a cut-off on bound states~\cite{BRODSKY1998299}. A particular way of doing this is achieved by only considering Fock states~$\sket{\{\widehat{p}_j,\widehat{\occ}_j\}}$ of invariant momentum squared below $\Qtransf^2$ in the expansion of the bound state~$\sket{\bound{\Psi}[][\Qtransf]}$.
In the absence of spin and transverse directions this constraint is:
\begin{equation}
    \label{pokeQ}
    P^+ P^-_{\text{free}} =
    \biggl( \sum_j \widehat{\occ}_j \widehat{p}^{\,+}_{j} \biggr)
    \biggl( \sum_j \widehat{\occ}_j \widehat{p}^{\,-}_{j} \biggr)
     = K
     \biggl(\sum_j \widehat{\occ}_j
    \dfrac{\mass_j^2}{\widehat{n}_j}
    \biggr)
    \leq \Qtransf^2
    \ ,
\end{equation}
where the sums go over all the excited parton modes.\footnote{Note that in equation~\eqref{pokeQ} for the $1+1D$ theory, the only dimensionful quantities on the LHS are the masses (but not the box size $L$). In higher dimensions, the LHS of eq.~\eqref{fockcutoff} also depends on~$\Lambda_\perp$ and~$L_\perp$.} As follows from eq.~\eqref{coords}, in the LF formalism the states of large equal-time momentum are those with either ${p^+\to\infty}$ or ${p^+\to 0}$. While the former option is automatically excluded in a block of fixed $K$, condition~\eqref{pokeQ} ensures that the light-front momenta are not too small.
\begin{figure}
\centering
\includegraphics[width=.8\textwidth]{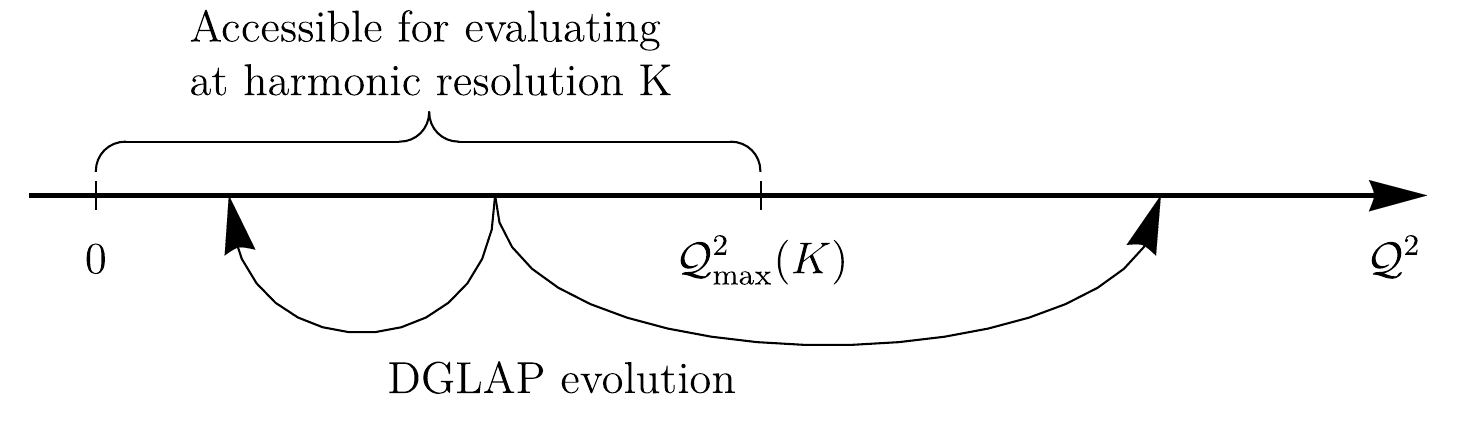}
\caption{At fixed harmonic resolution $K$, one can calculate PDFs up to the energy scale~$\Qtransf_{\max}^2(K)$. Once calculated at some energy scale, the PDFs can be evolved according to the Dokshitzer-Gribov–Lipatov–Altarelli–Parisi (DGLAP) equations.}
\label{fig:PDFevol}
\end{figure}
In terms of truncated bound states, one calculates the PDFs at probing scale $\Qtransf^2$ as:
\begin{equation}
    \label{pdfourcutoff}
    f_\ell (\Moden/K,\Qtransf)
    = \sand{\bound{\Psi}[][\Qtransf]}{\numberop_\ell}{\bound{\Psi}[][\Qtransf]} \ .
\end{equation}
This quantity is simply an expectation value of an unintegrated number operator, which may be calculated using a quantum computation that we discuss in the Sec.~\ref{algosec}.

\begin{figure}
\hspace{-1cm}
\centering
\includegraphics[width=.7\textwidth]{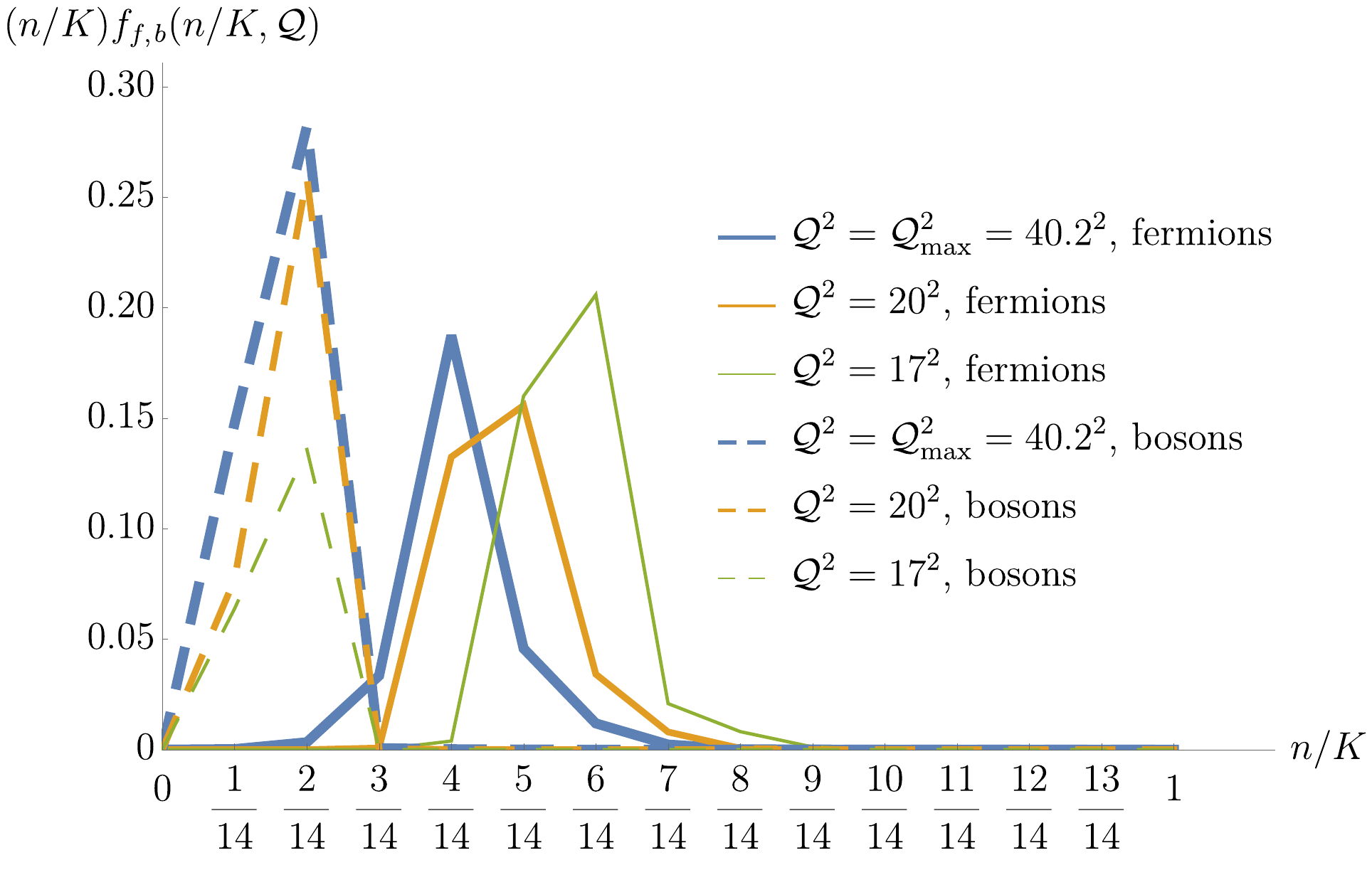}
\caption{Bosonic and fermionic parton distribution functions, as defined in eq.~\eqref{pdfour}, for the massive Yukawa model defined by~\eqref{Lagrangian} evaluated for harmonic resolution $K=14$. The values of parameters are chosen as in~\cite{pauli2}: $\widetilde{m}_B=6.7$, $\widetilde{m}_F=1$, $\lambda=1$, $\Lambda=2048$. Shown for the $M = 18.96$ eigenstate with different values of momentum cut-off: $\Qtransf^2=\Qtransf_{\max}^2,20^2,17^2$, where~$\Qtransf_{\max}^2=40.2^2$. The choice ${\Qtransf^2=\Qtransf_{\max}^2}$ corresponds to taking all the Fock states from the $K=14$ sector into account.}
\label{fig:pdfs}
\end{figure}

For a fixed $K$, there exists an upper bound on the \emph{free invariant mass} squared (the left-hand side of eq.~\eqref{pokeQ}). This sets the maximum energy scale~$\Qtransf_{\max}^2(K)$ up to which one can calculate PDFs at the given harmonic resolution. PDFs calculfvated at a particular value of~$\Qtransf^2$ can be evolved according to the Dokshitzer-Gribov–Lipatov–Altarelli–Parisi (DGLAP) equations~\cite{Gribov:1972ri,Altarelli:1977zs,Dokshitzer:1977sg,partondist,Collins:2011zzd,Peskin:1995ev}, including beyond~$\Qtransf_{\max}^2(K)$ if needed: this is illustrated in ~Fig.~\ref{fig:PDFevol}. We show bosonic and fermionic parton distribution functions for the massive Yukawa model defined in Sec.~\ref{themodelsection} evaluated at harmonic resolution $K=14$ in Fig.~\ref{fig:pdfs}.

\newpage
\section{Quantum Simulation in the Light Front\label{algosec}}

In this Section we present an algorithm for simulation of QFT in the front form on a digital quantum computer. In Sec.~\ref{hilbsize} we describe the scaling with harmonic resolution of the dimension of the Hilbert space. In Sec.~\ref{statesandops} we present three encodings, and explain the trade-offs between efficiency of encoding and simplicity of encoded operations. In Sec.~\ref{timeevol} we discuss the cost of time evolution. In Sec.~\ref{stateprep} we discuss the preparation of bound states~--- the eigenstates of the interacting Hamiltonian. In Sec.~\ref{measurement}, we discuss the measurement procedure used to obtain PDFS which, as was explained in Sec.~\ref{pdfs}, reduces in the DLCQ formalism to evaluating the expectation value of the number operator~\eqref{pdfourcutoff}. The methods we describe in this Section are optimal in both qubits and gates up to logarithmic factors.

In what follows all the resource requirements will be given in terms of the harmonic resolution $K$ (the dimensionless light-cone momentum), for two reasons. First, numerous quantities of physical interest can be calculated within a single Fock space sector with fixed~$K$. In \mbox{$1+1D$}, one example is given by the PDF, our focus in this paper; another example is the hadronic tensor~\cite{Lamm:2019uyc}. A straightforward generalization to \mbox{$3+1D$} would allow one to calculate electromagnetic form factors and decay constants: this is discussed in Sec.~\ref{qcdlf}. In such calculations, $K$ will define the number of points at which these quantities are evaluated. The calculation of dynamical quantities like cross-sections requires wave packets expanded in a basis of states having different total light-front momenta. In this case one would consider all the blocks of size up to some maximum $K$. In both cases $K$ controls the resolving power of the theory in the light-front, and so is a natural quantity with which to express computational cost.

\subsection{Hilbert space dimension \label{hilbsize}}

 In $1+1D$, for each harmonic resolution $K$ we have a finite-dimensional Hilbert space~$\hilbdimK$, which can be further split into blocks~$\hilbdimKQ$ of fixed charge $Q$. A lower bound on the dimension of $\hilbdimK$ is given by considering bosonic configurations only, which belong to the~$\hilbdim_{K,0}$ subblock. These are labeled by integer partitions of $K$, where the momenta $\widetilde{n}_j$ are the parts of the partition, and the occupancies $\widetilde{\occ}_j$ are the multiplicities:
\begin{equation}
    \label{bosonicpartitions}
    \{(\widetilde{n}_j,\widetilde{\occ}_j)|1\leq j\leq\widetilde{N}|\sum_j \widetilde{\occ}_j\widetilde{n}_j=K\}\ .
\end{equation}
The number of partitions of $K$ is denoted $p(K)$: its asymptotic behavior is ${\log_2 p(K) = \Theta(\sqrt{K})}$ (see for example Ch.~5 of~\cite{andrews1998theory}). Therefore
$\dim \hilbdimK\geq\dim \hilbdim_{K,0}\geq p(K)$.

Adding fermions and antifermions gives a subleading correction to the dimension of~$\hilbdimK$, since their occupancies are~$\occ_j,\overline{\occ}_j\in\{0,1\}$ due to the Pauli exclusion principle. Restricting to a particular value of~$Q$ does not change the aymptotic behavior either: for nonzero~$Q$ we have
\begin{equation}
    \dim \hilbdimKQ
    \geq
    \dim \hilbdim_{K-Q(Q+1)/2,0}
    \geq
    p\left(K-{Q(Q+1)}/{2}\right) \ ,
\end{equation}
since all purely bosonic configurations of the~$\hilbdim_{K-Q(Q+1)/2,0}$ sector can be turned into those of~$\hilbdimKQ$ by adding fermions with momenta ranging from~$0$ to~$Q$.
Therefore, independent of whether or not we restrict to a fixed value of~$Q$, the asymptotic behavior of the Hilbert space dimension is~$\Theta(\exp(\sqrt{K}))$.


\subsection{State encoding \label{statesandops}}

The light-front representation of the theory has given us a formulation in terms of orbitals representing fermions, antifermions or bosons with given momentum. We can use an analogue of the \emph{direct} mapping as in quantum chemistry~\cite{aspuru2005simulated,somma2002simulating}, which amounts to assigning a particular qubit register to each momentum mode. For fermions, this means having a single qubit for each fermionic degree of freedom. The anticommuting creation and annihilation operators can be defined with the aid of the Jordan-Wigner, Bravyi-Kitaev, or other related mappings~\cite{jordanwigner,bravyi2002fermionic,32JCP,BK2015,whitsuper,wehnerfermion}. The mapping of bosonic degrees of freedom has been previously studied in~\cite{somma2005quantum,macridin2018electron,mcardle2018quantum,sawaya2019quantum}. We consider two variations of the direct scheme which differ in how the bosons are encoded. The resulting encoding schemes will be referred to as the \emph{direct-direct} or \emph{direct-compact} mappings.

Within a block of harmonic resolution $K$, the occupancy $\widetilde{\occ}_\Moden$ of the bosonic mode of momentum $\Moden$ is bounded by the requirement that the maximum momentum carried by that mode is at most the total light-front momentum:~${\widetilde{\occ}_j \leq \maxoc_{\Moden}}$, where~${\maxoc_{\Moden} = \lfloor{K}/{{\Moden}}\rfloor}$.

The \emph{direct-direct} mapping, first introduced in~\cite{somma2005quantum}, uses a unary encoding requiring $\maxoc_{\Moden}$ qubits for storing $\maxoc_{\Moden}+1$ levels of each bosonic mode. This results in a total of $O(K\log  K)$ qubits. The bosonic creation and annihilation operators acting on the $\Moden$-th mode are represented by a sum of $\maxoc_{\Moden}$ \mbox{$2$-local} terms. However, since the locality of the fermionic operators is at least logarithmic in $K$, one may naturally want to trade locality of bosonic operators for a reduced number of qubits.

We therefore describe the \emph{direct-compact} mapping, which uses a binary encoding of the occupation number of the bosonic modes and requires $\lceil \log_2 \maxoc_{\Moden}\rceil$
qubits for encoding $\maxoc_{\Moden}$ levels, giving a total of $O(K)$ qubits. In this case, the creation and annihilation operators contain a sum of  $\maxoc_{\Moden}$ terms, each of which is \mbox{${\log_2 \maxoc_{\Moden}}$-local}. This encoding was recently used to describe molecular vibrations~\cite{mcardle2018quantum,macridin2018electron,sawaya2019quantum} and is described in detail in~\cite{mcardle2018quantum}. A related mapping is described in~\cite{macridin2018electron}.

\begin{figure}[ht]
	\centering
	\includegraphics[width=.8\textwidth]{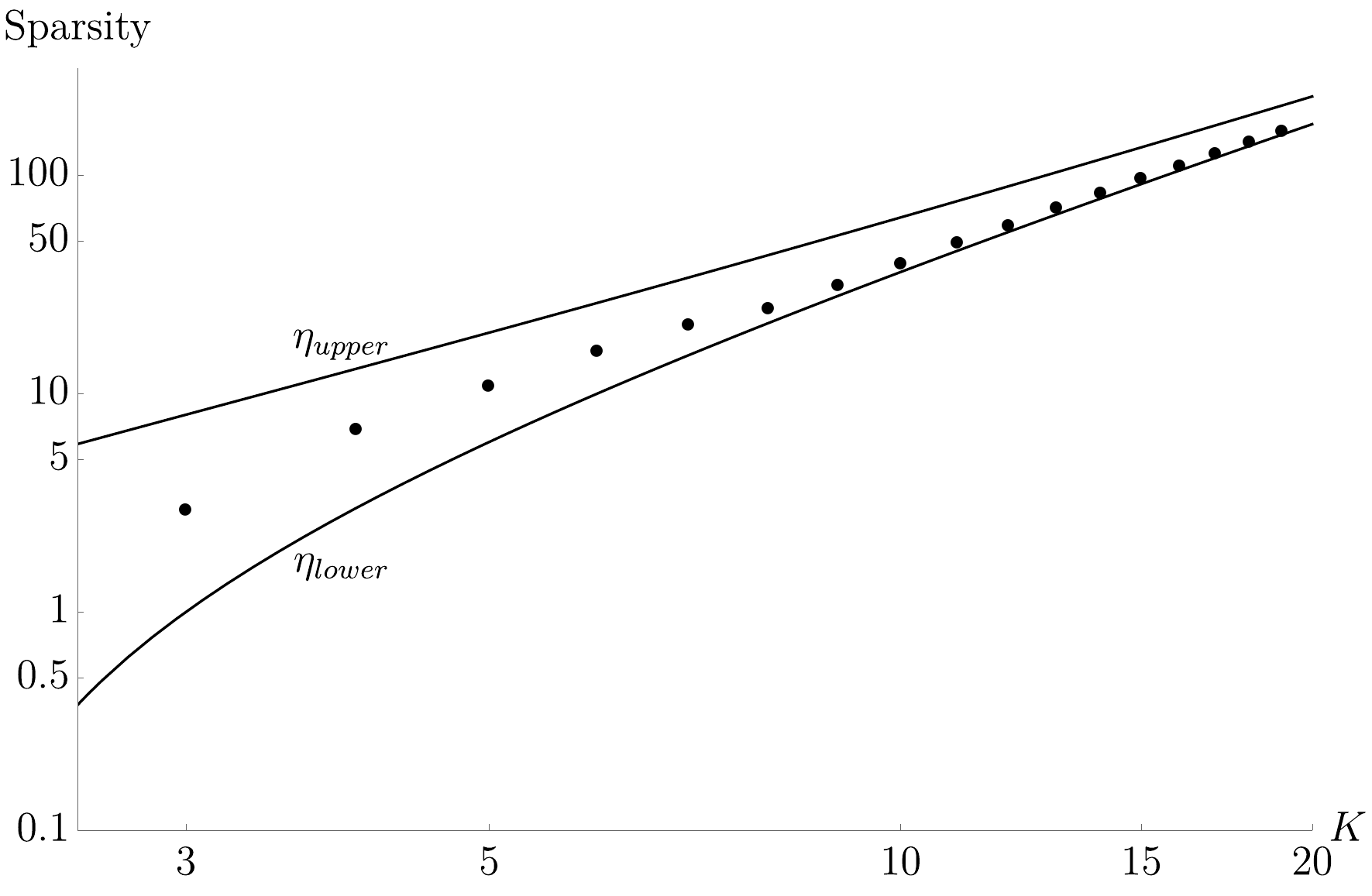}
	\caption{
	Hamiltonian sparsity vs. $K$. The curves label the upper and lower bounds on the sparsity, while the data points mark the exact sparsities for ${K=3,4,\ldots,19}$. The upper and lower bounds are given by
	${\spaM=\frac{1}{2}K^2+\frac{3}{2}K-1}$ and ${\spam=\frac{1}{2}K^2-\frac{3}{2}K+1}$ (derived in~App.~\ref{appsparse}).
	\label{sparsityfig}
	}
\end{figure}

The optimal encoding in terms of qubit resources is the \emph{compact} encoding scheme. This was first described for chemistry in~\cite{aspuru2005simulated} and efficient algorithms were given in~\cite{babbush2016exponentially, Toloui}.  For our model the compact mapping stores only the momentum modes with nonzero occupancies:
\begin{equation}
    \label{compactstate}
    |
    (\widehat{n}_{1},\widehat{\occ}_{1}),
    \,(\widehat{n}_{2},\widehat{\occ}_{2}),
    \ldots
    \rangle
    \ .
\end{equation}
For such an encoding, the number of qubits scales as $O(\sqrt{K}\log K)$; by comparing this to the Hilbert space dimension (see Sec.~\ref{hilbsize}) we can see that it is indeed optimal up to logarithmic factors. The compact encoding is discussed in detail in App.~\ref{partitions}. Use of this fully compact scheme requires simulation algorithms which depend on the sparsity of the Hamiltonian in the chosen basis. Fortunately, methods based on sparsity scale optimally with almost all simulation parameters~\cite{berry2015simulating,berry2017exponential}. The sparsity of the Hamiltonian of our model is shown in Fig.~\ref{sparsityfig} and discussed in detail in Sec.~\ref{timeevol}. We focus our efforts on quantifying the simulation complexity of the compact mapping because of its optimality. The properties of the three mappings are summarized in Table~\ref{tab:mapping}.

\def\arraystretch{1.4}
\begin{table}[h]
    \centering
    \begin{tabular}{|c|c|c|c|}

    \hline
    Mapping &
    \makecell{Qubit number, $\qubitn$} &
    \makecell{Hamiltonian\\locality} &
    \makecell{Hamiltonian\\sparsity} \\
    \hline
    Direct-Direct  & $O(K \log K)$       &$O(\log K)$         &   N/A \\         
    \hline
 Direct-Compact & $O(K)$
 &$O(\log K)$    &N/A  \\           
    \hline
    Compact     &$O(\sqrt{K}\log K)$      &  N/A             &$O(K^2)$ \\
    \hline
    \end{tabular}
    \caption{
    Dependence on $K$ of properties of the three encodings of the Fock space in ${1+1D}$. The direct mappings (which store the occupancies of all momentum modes) require $\widetilde{O}(K)$ qubits to encode $K$ modes. In these schemes the Hamiltonian is a sum of Pauli terms of locality ${O(\log K)}$.
    The compact mapping stores the occupancies of only nonempty momentum modes, giving an asymptotic scaling of $\widetilde{O}(\sqrt{K})$ qubits (which is \emph{optimal} up to logarithmic factors~--- see Sec.~\ref{hilbsize}). However, the Hamiltonian is no longer local in this encoding, so we must instead use sparsity-based techniques for simulation (discussed in Sec.~\ref{stateprep} and App.~\ref{appimplementation}).
    }
    \label{tab:mapping}
\end{table}

\subsection{Time evolution at constant harmonic resolution\label{timeevol}}

The goal of our simulation algorithm is first to prepare the eigenstates of the interacting quantum field theory described by Lagrangian given in eq.~\eqref{Lagrangian}. In each sector of fixed harmonic resolution~$K$ and charge~$Q$, the lowest mass-energy particle is a physical particle of the theory. We then aim to perform measurements on the state to determine properties of these composite particles such as PDFs and form factors.

State preparation is a basic element of any quantum simulation algorithm. In this section we give bounds on the cost in terms of quantum gates required to evolve a state in a subspace of fixed harmonic resolution $K$ for time $t$, to precision $\epsilon$. We use the methods of~\cite{berry2015simulating,berry2017exponential,berry2019time}, which are optimal in all relevant parameters.

Sparse Hamiltonians may be specified efficiently by two oracles: functions that can be called to give the defining information for the Hamiltonian. In App.~\ref{appimplementation} we give details of implementing two oracles needed by the methods of~\cite{berry2015simulating,berry2017exponential}. The first is $O_F$~--- an oracle that enumerates the positions of non-zero entries of the Hamiltonian in a given row. $O_F$~is defined in App.~\ref{enumerateH} where we show that the cost of~$O_F$ for the compact mapping is~$O(\sqrt{K}\log K)$.
The second is $O_H$, an oracle that computes the value of a nonzero entry to $\bitprec$ bits of precision given its indices. $O_H$~is defined in App.~\ref{computematrixelements} where we show that the cost of $O_H$ for the compact mapping is $O\bigl(K\log K + \bitprec^2\log \bitprec\bigr)$.

Using Theorem 1 from~\cite{berry2017exponential}, simulation of time evolution for time $t$ under a Hamiltonian on $n$ qubits of sparsity $d$ and maximum matrix element $||H||_{\rm max}$ to precision $\epsilon$ is given in terms of the parameter $\tau=d||H||_{\rm max}t$. The number of calls to $O_H$ and $O_F$ is
\begin{equation}
O\biggl(\tau\frac{\log\tau/\epsilon}{\log\log\tau/\epsilon}\biggr) \ ,
\end{equation}
and an additional
\begin{equation}
O\biggl(\tau[n+\log^{5/2}(\tau/\epsilon)]\frac{\log\tau/\epsilon}{\log\log\tau/\epsilon}\biggr)
\end{equation}
gates are required.

To simulate time evolution in a subspace of constant harmonic resolution $K$ for time $t$ in the compact mapping we have $n=O(\sqrt{K}\log K)$, $||H||_{\rm max}=O(K\log K/\Lambda)$, $d=O(K^2)$ and hence $\tau= O(tK^3\log K/\Lambda)$. The number of oracle calls required is then $\widetilde{O}(tK^3)$, and the number of gates required for this number of calls is $\widetilde{O}(tK^4)$ if $\bitprec$ is polylogarithmic in $K$. The number of additional gates required is $\widetilde{O}(tK^{7/2})$ and so the overall simulation cost up to logarithmic factors is $\widetilde{O}(tK^4)$.

\subsection{State preparation\label{stateprep}}

State preparation by any of the standard schemes requires simulation of time evolution. These schemes include phase estimation, adiabatic state preparation ~\cite{farhi2000,Aharonov:2003:AQS:780542.780546,preskill1} as well as variational approaches~\cite{peruzzo2014variational} and quantum imaginary time evolution~\cite{motta2019determining}. Adiabatic state preparation performs time dependent evolution under a Hamiltonian varying along a path connecting the target Hamiltonian to some initial, simple Hamiltonian. The minimum time this evolution can take while preserving the system in the ground state is determined by the minimum gap along the path. To determine the cost of adiabatic state preparation we must bound the spectral gap along a chosen adiabatic path, either rigorously or by invoking physical arguments. Assuming a gapped adiabatic path can be found, one must quantify the cost of simulation of evolution under a time-dependent Hamiltonian to perform adiabatic state preparation.

In our system, $K$ controls the precision with which the theory describes the field theory in the front-form. We consider adiabatic paths such that the max norm of the Hamiltonian is everywhere upper bounded by the max norm of the Hamiltonian with the final $K$. We conjecture that amongst such paths a gapped adiabatic path exists that connects the theory at low $K$ to the theory at high $K$. The property of the space of paths that we shall use in the analysis below is that the max norm of the Hamiltonian varies as ${O(tK/T)}$ for ${t\in[0,T]}$, where $T$ is the length of the adiabatic evolution.

We will use the results of~\cite{berry2019time} to bound the cost of adiabatic state preparation. Specifically, we use Theorem 10 of~\cite{berry2019time} which, given a Hamiltonian on $n$ qubits, with sparsity $d$ and a bound on the integral of the maximum matrix element of $||H||_{\rm max}$ along the path, $||H||_{\rm max,1}$, gives the number of queries to $H_F$ and $H_O$ required as:
\begin{equation}
O\biggl(d||H||_{\rm max}\frac{\log d||H||_{\rm max,1}/\epsilon }{\log\log d||H||_{\rm max,1}/\epsilon}\biggr)
\end{equation}
and a number of additional gates scaling as $\widetilde{O}(d ||H||_{\rm max,1} n)$. This method requires two additional oracles: one to compute a scaling factor, and one to compute the time-dependent max norm. Many paths obeying our max norm condition can by realized with $O(\log K)$ gates and so only change the scaling by logarithmic factors.

The cost of adiabatic evolution for time $T$ by this method is given by setting $d=\widetilde{O}(K^2)$,  $||H||_{\rm max,1}=KT$, and using the costs of the oracles $O_H$ and $O_F$ above to obtain a total cost of $\widetilde{O}(K^4T)$, which is the same as that for time-independent simulation. It remains a matter of future work to provide rigorous results on the efficiency of this, and other, adiabatic state preparation procedures.

Our Hamiltonian commutes with both $K$ and $Q$, and we are interested in preparing states of specific charge in a sector of fixed $K$. Our compact encodings do not restrict to states of fixed charge. Exact evolution under the Hamiltonian will preserve the expectation values of these quantities given by the initial state. However, time dependent evolution or approximate evolution under the Hamiltonian may cause leakage to states of different $K$ and $Q$ for the direct mappings, and states of different $Q$ in the case of compact mappings. We can therefore improve our state preparation by using phase estimation of $K$ and $Q$ after adiabatic evolution to project back to the desired sector. If the leakage to sectors of incorrect $K$ or $Q$ is small, then with high probability phase estimation of those operators will project us to the desired value. In the low-probability case that phase estimation results in the incorrect value of $K$ or~$Q$, we simply discard the result and start the state preparation procedure again.

The existence of small example problems for small $K$ makes the implementation of such calculations on Noisy Intermediate Scale Quantum (NISQ) computers a possibility. Such calculations would attempt to variationally minimize the expectation of the invariant mass in a given sector of $K$ and $Q$, and then perform measurements to estimate the PDFs in this variational ansatz~\cite{peruzzo2014variational}. An alternative to variational optimization of an ansatz is to use another heuristic such as quantum imaginary time evolution (QITE)~\cite{motta2019determining,mcardle2019variational}. The BLFQ formulation~\cite{varybasis} is particularly useful when one considers NISQ implementations~\cite{VQEBLFQ}.

\subsection{Measurement\label{measurement}}

One of the benefits of the LF formulation of QFT for quantum simulation is the simple form of measurement operators.  This is due to the fact that one can calculate various observables directly from the LF wave function~\cite{BRODSKY1998299}. The determination of PDFs values at a fixed total light-front momentum $K$ can be accomplished by estimation of single-mode operators,as can be seen from eq.~\eqref{pdfour}.\footnote{As we mentioned above, in a more general case one needs to measure the sum of single-mode operators over the transverse directions and additional quantum numbers, \mbox{eqs.~\eqref{pdfgen}-\eqref{numberoperator}} below. This only changes the complexity polynomially.} This task can be performed efficiently on a quantum computer.

For the direct mappings, eq.~\eqref{numberoperatorsour} could be written simply as a sum over projectors on the qubit registers corresponding to the occupancies of modes of momentum $\Moden$. For the compact mappings, we also wish to construct an operator whose eigenstates are the compact Fock states, and whose eigenvalue for a particular Fock state is the occupation of a particular mode $\Moden$. The task is more complicated for the compact encoding, because a particular momentum mode is not always encoded in the same register of qubits, but it is still efficiently tractable. In brief, if we wish to extract the occupancy of a momentum mode $\Moden$, then on each register that encodes a mode $n_{j}$ and its occupation $\occ_j$ we must perform a locally-controlled operation that adds $\occ_j$ to a fixed ancillary register if and only if $n_{j}=\Moden$. After performing this task on the fermions, antifermions, and bosons, the ancillary register will encode the total occupation of $\Moden$, which we can then simply read out.
This method is efficient. By employing a slightly more complicated scheme, we can further improve the efficiency of this operation; details are given in App.~\ref{occupationmeasurement}, and the resulting number of CNOTs and single-qubit operations required is $\widetilde{O}(\sqrt{K}+\bitprec)$ for $\bitprec$ bits of precision.

The dependence of the PDF on the probing scale~$\Qtransf^2$ is introduced by imposing a momentum cut-off as in equation~\eqref{pokeQ}. This is also easily accomplished within the second-quantized formalism. Indeed, since the Fock states are the eigenstates of the free Hamiltonian, calculating the quantity on the LHS of~\eqref{pokeQ} can be achieved by running the phase estimation algorithm for the free Hamiltonian. It only remains to introduce an ancillary register storing the information on whether the particular Fock state has to be kept in the expansion.


\subsection{Mass renormalization\label{massrenorm}}

Up to this point we have not discussed the issue of mass renormalization in our model, which was studied in detail in~\cite{pauli2}. Thus, we have implicitly assumed that the parameters in the Lagrangian are also those which would be measured in a thought experiment. This is only approximately correct for weak couplings and fails in the strong coupling regime. In order to correctly determine the eigenvalues and eigenstates of the $M^2$ operator for arbitrary values of coupling we must proceed as follows. Given as input the finite \emph{renormalized} (i.e., physically observed) masses $\widetilde{\mass}_B$ and $\widetilde{\mass}_F$ and the coupling $\lambda$, we first determine the values of the \emph{bare} constants $\mass_B$ and $\mass_F$ appearing in the Lagrangian. At a given harmonic resolution $K$, this is achieved by varying the bare masses $\mass_B$, $\mass_F$ for fixed $\lambda$ to satisfy the condition:
\begin{equation}\label{renormcondition}
    \begin{split}
    \lowe
    \left\{
    M^2_{K,Q=0} (\mass_B, \mass_F, \lambda, \Lambda)
    \right\}
    = \widetilde{\mass}_B \ ,
    \\
    \lowe
    \left\{
    M^2_{K,Q=1} (\mass_B, \mass_F, \lambda, \Lambda)
    \right\}
    = \widetilde{\mass}_F \ ,
    \end{split}
\end{equation}
which implies that the lowest eigenvalues in the ${Q=0}$ and $Q=1$ sectors of the mass matrix are associated with the physical boson and fermion masses. Having thus determined the bare couplings $\mass_B$ and $\mass_F$, the $M^2$ operator now reproduces the spectrum of the theory at harmonic resolution $K$.

To solve~\eqref{renormcondition} one performs a gradient search in the two-dimensional space $(\mass_B, \mass_F)$~\cite{pauli2}.
The starting value $\mass_B^{(0)}$ can be analytically calculated from the $K=2$ sector of the model:
\begin{equation}
     \mass_B^2 = \widetilde{\mass}_B^2 - \dfrac{\alpha_2}{4\pi} \lambda^2
 \ ,
\end{equation}
where $\alpha_2$ is a function of $\Lambda$ defined as in~\eqref{intertias} below.
The starting value $\mass_F^{(0)}$ is then found by substituting $\mass_B^{(0)}$ into the second line of~\eqref{renormcondition} and performing gradient search in $\mass_F$:
\begin{equation}
     \lowe
\left\{
M^2_{K,Q=1} (\mass_B^{(0)}, \mass_F, \lambda, \Lambda)
\right\}
= \widetilde{\mass}_F \ .
\end{equation}
Each iteration of the gradient search corresponds to a run of an algorithm described in the previous section. When conditions (18) are satisfied with the desired precision, the wave functions can be used for calculating the PDFs.

The renormalizability of the theory guarantees the convergence of the method in the~${\Lambda\to\infty}$ limit: the physical masses do not depend on the cutoff~$\Lambda$~\cite{pauli2}. Moreover, the fact that the Hamiltonian norm only depends on~$\Lambda$ logarithmically (see~App.~\ref{appnorm}) means that choosing sufficiently large~$\Lambda$
to obtain convergence does not require exorbitant resources.

If one moves further to calculating dynamic quantities, such as cross-sections, one would similarly have to perform the renormalization of the coupling constant $\lambda$, which would amount to performing a gradient search in the three-dimensional space of bare couplings with an additional condition on a certain amplitude.\footnote{Strictly speaking, since the coupling constant also has to be determined from an experiment (similar to the masses), one needs to implement this procedure even to calculate static quantities. Calculating cross-sections amounts to expanding the wave packets within Hamiltonian blocks of different $K$. However, due to the exponential growth of the Hilbert space size with $\sqrt{K}$, the qubit asymptotics will remain unchanged.
}


\section{\emph{Ab Initio} Simulation of QCD in the Light Front\label{qcdlf}}

We now discuss how the machinery developed in Secs.~\ref{themodelsection} and~\ref{algosec} can be generalized to QCD in~$3+1D$. We briefly review the notations of QCD, discuss the qubit requirements, and present the expressions for observables in a form suitable for quantum simulation.
We will see that our method gives asymptotic improvement in the scaling of qubit resources with cutoffs over previous simulation methods based on equal time formulation~\cite{preskill2018simulating}. This results in several orders of magnitude fewer qubits for the smallest physically-meaningful cutoffs~\cite{Lamm:2019uyc}.

\subsection{Light-front QCD in $3+1D$}

\begin{sloppypar}
QCD is a field theory of Dirac fermions (quarks) interacting via an $SU(3)$ `color' gauge field. Due to the non-abelian nature of the gauge group, the mediators of the color interaction (gluons) carry the color charges themselves and can directly interact with each other. The fermionic field $\Psi_{\ccolor,\alpha}$  transforms under the fundamental representation of the gauge group, ${\ccolor=1,2,3}$ being the index in the color space, and ${\alpha=1,2,3,4}$ being the Dirac spinor index. The spinor $\Psi_{\ccolor,\alpha}$ of color $\ccolor$ and its adjoint ${\overline{\Psi}_{\ccolor,\alpha}=\Psi^\dagger_{\ccolor,\beta}(\gamma^0)_{\beta\alpha}}$ each have four complex components. The gauge field vector potentials $\bm{A}^\mu\!\!{}_{\ccolor\ccolor\mathrlap{{}^{\prime}}\mathstrut}$ transform under the adjoint representation of the gauge group, and can be expanded as ${\bm{A}^\mu\!\!{}_{\ccolor\ccolor\mathrlap{{}^{\prime}}\mathstrut} = A^\mu\!\!{}_{\acolor\vphantom{{}^{\prime}}\mathstrut} T^\acolor\!\!\!{}_{\ccolor\ccolor\mathrlap{{}^{\prime}}\mathstrut}}\,$, where $A^\mu\!\!{}_{\acolor\vphantom{{}^{\prime}}}$ are the eight real vector fields, while $T^\acolor_{\ccolor\ccolor'}$ are the generators of the gauge group obeying\s\footnote{Note that one only has to be careful when raising and lowering spacetime indices because their metric is non-trivial.
}
\end{sloppypar}
\begin{equation}
    \label{su3algebra}
    [T^\rcolor,\,T^\scolor]_{\ccolor\ccolor'\mathstrut} = \iu f^{\rcolor\scolor\acolor} T^\acolor\vphantom{]}_{\ccolor\ccolor'\mathstrut}
     \ ,\qquad
    \Tr(T^\acolor T^\bcolor) = \dfrac{1}{2} \delta^{\acolor\bcolor}
     \ ,
\end{equation}
$f^{\rcolor\scolor\acolor}$ being the structure constants of the $\mathfrak{su}(3)$ algebra.

For $N_f$ flavors of quarks, the QCD Lagrangian acquires the following form:
\begin{equation}
    \label{lqcd}
    \mathcal{L} = -\dfrac{1}{4} G^{\mu\nu}_a G^a_{\mu\nu} +
    \sum \limits_{\Modeflavor=1}^{N_\Modeflavor}
    \dfrac{1}{2}
    \Bigl[
    \overline{\Psi}_{\ccolor,\alpha}^{\Modeflavor} (\iu \gamma_\mu^{\alpha\beta} \bm{D}^\mu\!\!\!{}_{\ccolor\ccolor'} - \mass^{\Modeflavor} \delta_{\ccolor\ccolor'}) \Psi^{\Modeflavor}\!\!{}_{\ccolor',\beta} + \text{h.c.}
    \Bigr]
     \ ,
\end{equation}
where
\begin{equation}
    \label{cemft}
    G^{\mu\nu}_\acolor = \partial^\mu A^\nu_\acolor - \partial^\nu A^\mu_\acolor - g f^{\acolor\rcolor\scolor} A^\mu_\rcolor A^\nu_\scolor
\end{equation}
is the color-electromagnetic field tensor, $\mass^\Modeflavor$ are the masses of different quark flavors, and
\begin{equation}
    \label{covd}
    \bm{D}^\mu\!\!\!{}_{\ccolor\ccolor\mathrlap{{}^{\prime}}\mathstrut} = \delta_{\ccolor\ccolor\mathrlap{{}^{\prime}}\mathstrut}\; \partial^\mu + \iu g \bm{A}^\mu\!\!{}_{\ccolor\ccolor\mathrlap{{}^{\prime}}\mathstrut}
\end{equation} is the covariant derivative.

In DLCQ, we shall use the collective label~$\Modecollective$ containing the following degrees of freedom for gluons and quarks:
\begin{subequations}
\label{labels}
\begin{alignat}{9}
    \label{labelB}
    \Modecollective &=
    \{\Moden,\vec{\Moden}_\perp,\helicity,\acolor \}
     \
    &&\text{(gluons)}&&,
\\
    \label{labelF}
    \Modecollective &=
    \{\Moden,\vec{\Moden}_\perp,\helicity,\ccolor,\Modeflavor\}
     \
    &&\text{(quarks)}&&,
\end{alignat}
\end{subequations}
where $\acolor$ is the color index in the adjoint representation, $\ccolor$ is the color index in the fundamental representation, $\Modeflavor$ is the flavor index, and $\helicity$ is polarization or helicity. The discretized light-front momentum ${n = 2\pi k_z/L}$ is analogous to that in \mbox{$1+1D$,} while ${{\vec{n}_\perp=(n_{x},n_y})}$ is the dimensionless internal momentum, defined by ${\vec{k}_{\perp}=(k_x,k_y)=2\pi\vec{n}_\perp/L_{\perp}}$, which is introduced in order to separate the center-of-mass motion of the composite state. For a Fock state~$\sket{\{\collective_j,\occ_j\}}$,
\begin{equation}
    \label{intrinsic}
    {\vec{k}_{\perp\,j}=\vec{p}_{\perp\,j}-x_j\vec{P}_\perp}
     \ ,\qquad
    \sum_j \occ_j \vec{k}_{\perp\,j}
    = \dfrac{2\pi}{L_{\mathrlap{\perp}}}\sum_j \occ_j \vec{n}_{\perp\,j}
    = 0
     \ ,
\end{equation}
where the sum goes over all the partons.

\begin{sloppypar}
In $3+1D$, one immediately benefits from using the light-front formulation of non-Abelian gauge theories because of the vacuum triviality and the absence of ghost fields~\cite{BRODSKY1998299}. However, the presence of the transverse directions necessitates an additional momentum cut-off~$\Lambda_\perp$.  The Hamiltonian matrix remains sparse~\cite{VaryAbInitio}, allowing one to use the algorithms discussed above.

Furthermore, in the DLCQ all the momentum modes, including those of massless bosons, necessarily carry a nonzero light-front momentum~\cite{BRODSKY1998299}, i.e., ${\Moden>0}$ in eq.~\eqref{labelB}.\footnote{The light-front zero mode $a_0$ requires special treatment; in particular, it carries the information about the equal-time vacuum of the theory ~\cite{BRODSKY1998299,Kalloniatis:1993jh,Kalloniatis:1994ei,Kalloniatis:1994fk,Kalloniatis:1994ix,Kalloniatis:1995fc,Pauli:1995dt}. By imposing antiperiodic boundary conditions on the LF fields, one may by able to completely eliminate the effect of zero modes~\cite{PhysRevD.66.045019}.}
Hence, although the qubit requirements in arbitrary dimension increase relative to the $1+1D$ case, their scaling with harmonic resolution $K$ only increases to $\widetilde{O}(K)$, since in the worst case the state may be composed of $K$ modes with light-front momentum one, all having distinct transverse momenta. Note that using the compact encoding (in the sense of only storing the occupied modes) is crucial: the number of unoccupied modes scales as the product of the momentum cutoffs over all dimensions.\footnote{Technically, the scaling including dimension in the light-front formulation is $\widetilde{O}(dK)$, whereas the scaling including dimension in equal-time is $\widetilde{O}(K\Lambda_\perp^{d-1})$. The factor of $d$ in the light-front scaling is due to the necessity of encoding the value of each component of momentum for each occupied mode. However, for a fixed theory, $d$ is a constant, so we may ignore it in the light-front scaling.}

In order to simulate the full QCD Lagrangian with harmonic resolution $K$ and transverse momentum cut-off $\Lambda_\perp$, an upper bound on the number of required qubits to store the light-front wavefunction for QCD in $3+1$ dimensions is:
\vspace{-.55cm}
\end{sloppypar}
\begin{equation}
    \label{qcdestimate}
    \begin{multlined}
    \qubitn \le
    \overbrace{
    \underbrace{
    2K
    \vphantom{n_f}
    }_{
    \llap{\hspace{.6cm}
    \parbox{2cm}{\centering \scriptsize
    \text{number of}
    \\
    \text{occupied}
    \\
    \text{fermion/antifermion}
    \\
    \text{modes}
    }
    }
    }
    \ \Bigl[
    \underbrace{\lceil\log_2K\rceil+2\lceil\log_2\Lambda_\perp\rceil}_{\text{momentum}}+\underbrace{1}_{\text{helicity}}+\underbrace{\lceil\log_2n_f\rceil}_{\text{flavors}}+\underbrace{\lceil\log_2n_c\rceil}_{\text{colors}}
    \Bigr]
    }^{\text{fermion/antifermion modes}}
    \vspace{-.6cm}
    \\
    \hspace{2.8cm}
    +
    \overbrace{
    \underbrace{
    K
    \vphantom{n_f}
    }_{
    \llap{\hspace{.6cm}
    \parbox{2cm}{\centering \scriptsize
    \text{number of}
    \\
    \text{occupied}
    \\
    \text{boson modes}
    }
    }
    }
    \Bigl[
    \underbrace{\lceil\log_2K\rceil+2\lceil\log_2\Lambda_\perp\rceil}_{\text{momentum}}+\underbrace{\lceil\log_2K\rceil}_{\text{occupancy}}+\underbrace{1}_{\text{helicity}}+\underbrace{\lceil\log_2(n_c^2-1)\rceil}_{\text{colors}}
    \Bigr]
    }^{\text{boson modes}}
    \ ,
    \vspace{0.2cm}
    \end{multlined}
\end{equation}
(see App.~\ref{appimplementation} for a more detailed analysis in the $1+1D$ case). The helicity is encoded by a single qubit because the LF Dirac spinor has two `good' (independent) components~\cite{Tang:1991rc}. The number of flavors~$n_f$ taken into consideration depends on the probing scale~$\Qtransf^2$. Evaluation of eqn.~\ref{qcdestimate} for the computation of~\cite{Lamm:2019uyc}, which requires $400000$ qubits for an equal time calculation on a $20^3$ lattice, yields $\qubitn=1360$ qubits (after additionally including ancillas required for the computations~--- see App.~\ref{appimplementation}). This reduction in qubit numbers will become even more dramatic with increasing lattice size and cutoffs.


\subsection{Parton distribution functions\label{pdfsinlf}}

\begin{sloppypar*}
In QCD, all the information about the hadronic part of a scattering process is encoded within the so-called \emph{hadronic tensor} $W_{\mu\nu}$, also known as the \emph{forward Compton amplitude}~\cite{Peskin:1995ev},
\begin{equation}
    \label{hadronictensor}
        W_{\mu\nu} = \iu \int \d{}^4x\,\me^{\iu qx} \sand{P}{T\{[J_\mu^\dagger(x)J_\nu(0)\}}{P}  \ ,
\end{equation}
where~$\sket{P}$ is a hadronic state of four-momentum $P$ (averaging over spins is implied unless~$\sket{P}$ is spinless), and ${J_\mu(x) = \sum_\Modeflavor \charge_\Modeflavor \overline{\Psi}_\Modeflavor(x) \gamma_\mu \Psi_\Modeflavor(x)}$ is the quark current operator ($\charge_\Modeflavor$~being the quark charges; the sum is taken over all the quark flavors).
\end{sloppypar*}

In the case of deep inelastic scattering
${l(k) + p(P) \to l(k-q) + X(P+q)}$,
where $l$ is the lepton of momentum $k$, $p$ is the proton of momentum $P$, and $X$ is the final hadronic state of momentum $P+q$, one can write $W_{\mu\nu}$ in terms of two scalar \emph{structure functions} $W_{1,2}$ as
\begin{equation}
    \label{hadronictensordis}
        W_{\mu\nu}
        =
        \left(-g^{\mu\nu}+\dfrac{q^\mu q^\nu}{q^2}\right)W_1(x,\Qtransf^2)+\left(P^\mu-q^\mu\dfrac{P\cdot q}{q^2}\right)\left(P^\nu-q^\nu\dfrac{P\cdot q}{q^2}\right)W_2(x,\Qtransf^2)  \ ,
\end{equation}
where $q^2=-\Qtransf^2$,
and ${x=\Qtransf^2/(2P\cdot q)}$~\cite{Peskin:1995ev}. According to the optical theorem, the cross-section of such an inclusive process is given by the imaginary part of~$W_{\mu\nu}$, and hence of $W_1$ and $W_2$.

Within the parton model (i.e., to the zeroth order in the strong coupling constant~$\alpha_S$),  $\Im W_1$ and $\Im W_2$ can be expressed as
\begin{equation}
    \label{Wpdfs}
    \Im W_1 (x,\Qtransf^2) = \dfrac{P\cdot q}{2x} \Im W_2(x,\Qtransf^2) = \pi \sum_\Modeflavor \charge_\Modeflavor^2 f_\Modeflavor(x,\Qtransf^2)  \ ,
\end{equation}
where
$f_\Modeflavor(x,\Qtransf^2)$ are the \emph{parton distribution functions} (PDFs),
and the sum is taken over all the quark flavors contributing at the energy scale~$\Qtransf^2$.\footnote{For example, ${n_f=3}$ at ${\Qtransf^2 = (1\mathrm{GeV})^2}$ and ${n_f=5}$ at $\Qtransf^2 =(90 \mathrm{GeV})^2$.
}

In the light-front formalism, PDFs represent the probability of finding a parton of a given type carrying the longitudinal momentum fraction
\begin{equation}
    x = \dfrac{\Modep^{\mathrlap{+}}}{\bsm
    } = \dfrac{n}{K}
     \ ,\qquad
    0<x\leq1
\end{equation}
inside a hadron of a total longitudinal momentum $P^+$.
Formally, quark PDFs are defined as matrix elements of the quark field operators\footnote{Eqs.~\eqref{quarkpdf} and~\eqref{gluonpdf} are written in the light-cone gauge~$A^+_\acolor=0$, and, therefore, do not contain the Wilson line operator~\cite{Collins:2011zzd}.
}~\cite{Collins:2011zzd}:
\begin{equation} 
    \label{quarkpdf}
    f_\Modeflavor(x) =
    \sum_{\helicity,\ccolor,\Modeflavor}
    \int \dfrac{\d{}\Modep^-}{4\pi} \, \me^{- \iu x P^+ \Modep^-} \sand{P}{\overline{\psi}_{\helicity,\ccolor,\Modeflavor}(0^+,\Modep^-,\vec{0}_\perp)\gamma^+\psi_{\helicity,\ccolor,\Modeflavor}(0)}{P} \ ,
\end{equation}
where we suppress the $\Qtransf^2$-dependence and assume that ${\sbraket{P}{P}=1}$.

The gluon PDF emerges as one further evaluates eq.~\eqref{Wpdfs} to the first order in~$\alpha_S$. It is defined as
\begin{equation}
    \label{gluonpdf}
    f_g(x) =
    \sum_\acolor
    \int \dfrac{\d{}\Modep^-}{2\pi xP^+} \, \me^{- \iu x P^+ \Modep^-}
    \sand{P}{G^{+i}_{\,\acolor}(0^+,\Modep^-,\vec{0}_\perp) G_{+i}^{\,\acolor}(0)}{P}
     \ .
\end{equation}
The normalization of PDFs is given by
    \begin{alignat}{9}
    \label{pdfnormalization}
    &\int_0^1 \d x\,x\biggl[\sum_\Modeflavor f_\Modeflavor(x)+f_g(x)\biggl] &&= 1  \ ,\qquad
    &\int_0^1 \d x\,\sum_\Modeflavor \charge_\Modeflavor f_\Modeflavor(x) &&= Q  \ ,
    \end{alignat}
which reflects the fact that that the individual momenta and charges of partons sum up to those of the hadron.

Upon substituting the LF free field expansion,~\eqref{quarkpdf} and~\eqref{gluonpdf} acquire a simple form in terms of the
number operators~\cite{Bouchiat:1971mj,Soper:1976jc,Collins:2011zzd}:
\begin{equation}
\begin{alignedat}{88}
    \label{pdfgen}
    &f_\Modeflavor(x) &&=
    &&\sand{P}{\numberop_\Modeflavor(x)}{P}
     \ &&,\qquad
    &f_g(x) &&=
    &&\sand{P}{\numberop_{\mathrlap{g}\hphantom{\Modeflavor}}(x)}{P}
     \ &&,
\end{alignedat}
\end{equation}
where
\begin{equation}
\begin{alignedat}{10}
    \label{numberoperator}
    &\numberop_\Modeflavor(x)
    &&= \numberop_\Modeflavor(\Moden/K)
    &&= \sum_{\helicity,\ccolor,\vec{\Moden}_{\mathrlap\perp}}
    \OPbd_\Modecollective
    \OPb_\Modecollective
     \ &&,\qquad
    &\numberop_g(x)
    &&= \numberop_g(\Moden/K)
    &&= \sum_{\helicity,\acolor,\vec{\Moden}_{\mathrlap\perp}}
    \OPad_\Modecollective
    \OPa_\Modecollective
     \ &&.
\end{alignedat}
\end{equation}
\begin{sloppypar*}
In~\eqref{numberoperator}, the operator $\OPbd_\Modecollective$ creates a quark with quantum numbers~${\Modecollective=\{\Moden,\vec{\Moden}_\perp,\helicity,\ccolor,\Modeflavor\}}$, while~$\OPad_\Modecollective$ creates a gluon with quantum numbers~${\Modecollective=\{\Moden,\vec{\Moden}_\perp,\helicity,\acolor\}}$.
\end{sloppypar*}

As in $1+1D$, within the discretized light-front formalism, introducing the momentum transfer $\Qtransf^2$ amounts to cutting off the total four-momentum of the Fock states in the expansion of the hadronic state~\cite{BRODSKY1998299}:
\begin{equation}
    f_\Modeflavor(x,\Qtransf^2) =
    \sand{P^{(\Qtransf)}}{\numberop_\Modeflavor}{P^{(\Qtransf)}}
     \ ,\qquad
    f_g(x,\Qtransf^2) =
    \sand{P^{(\Qtransf)}}{\numberop_g}{P^{(\Qtransf)}}
     \ .
\end{equation}

For a Fock state~$\sket{\{\collective_j,\occ_j\}}$ whose total four-momentum is given by
\begin{equation}
    \label{fockfourmomentum}
    \begin{gathered}
    P^+
    =  \sum_j \occ_j p^{+}_{j}
    = \dfrac{2\pi}{L} \sum_j \occ_j n_{j}
    = \dfrac{2\pi}{L} K
     \ ,\qquad
    \vec{P}_\perp
    =  \sum_j \occ_j \vec{p}_{\perp\,j}
    =  \dfrac{2\pi}{L_\perp}\sum_j \occ_j \vec{n}_{\perp\,j}
     \ ,\\
    P^-_{\text{free}} = \sum_j \occ_j \biggl( \dfrac{\mass_j^2 + \vec{p}_{\perp\,j}^{\,2}}{p^+_j} \biggr)
     \ ,
    \end{gathered}
\end{equation}
a particular Lorentz-invariant cut-off is provided by only considering Fock states of total invariant mass below $\Qtransf^2$~\cite{BRODSKY1998299}:
\begin{equation}
    \label{fockcutoff}
    P^+ P^-_{\text{free}} - (\vec{P}_\perp)^2
    = \sum_j \occ_j \biggl( \dfrac{\mass_j^2 + \vec{k}_{\perp\,j}^{\,2}}{x_j} \biggr)
    \leq \Qtransf^2
     \ ,
\end{equation}
where $\mass_j$, ${x_j=n_j/K}$ and $\occ_j$ are the parton masses, light-front momentum fractions and occupancies, respectively; the intrinsic momenta~$\vec{k}_\perp$ are defined as in~\eqref{intrinsic}, and the sum goes over all the occupied parton modes.

As we discussed in Sec.~\ref{pdfs} and illustrated in Fig.~\ref{fig:PDFevol}, for fixed harmonic resolution (and transverse cutoff in $3+1D$) the left-hand side of eq.~\eqref{fockcutoff} is bounded above by some energy scale ${\Qtransf^2_{\max}(K,\Lambda_\perp)}$. Once calculated at some scale ${\Qtransf^2\leq\Qtransf^2_{\max}(K,\Lambda_\perp)}$, the PDFs can be evolved according to the DGLAP equations~\cite{Gribov:1972ri,Altarelli:1977zs,Dokshitzer:1977sg,partondist,Collins:2011zzd,Peskin:1995ev}.

Expression~\eqref{numberoperator} is appealing from the quantum computational perspective because the number operator can be measured efficiently~--- see App.~\ref{occupationmeasurement}. This remains true if one wants to exclude certain Fock states from consideration according to~\eqref{fockcutoff}. In App.~\ref{appimplementation} we illustrate this by providing an explicit realization of these measurements for the model~\eqref{Lagrangian}.

As we mentioned above, the RHS of eq.~\eqref{Wpdfs} is the zero-order term in the perturbative expansion of the hadronic tensor in the powers of the strong coupling constant. It is obtained from eq.~\eqref{hadronictensor} by replacing the full Heisenberg currents~$J_\mu(x)$ with the currents~$j_\mu(x)$ of the non-interacting theory. Within the traditional approach, one calculates the hadronic tensor by obtaining higher-order perturbative corrections to eq.~\eqref{Wpdfs}. The paradigm of quantum simulation naturally suggests a different way to proceed: by switching to the interaction picture, we can move all of the complexity of the interacting theory into the state preparation stage. Doing so will have a minor effect on computational resources while allowing us to keep the measurement operators unchanged. Most importantly, such a calculation would be non-perturbative.


\subsection{Form factors and decay constants\label{appendixformfactors}}


In this subsection we derive expressions for the electromagnetic form factor of a hadronic state (similar to~\eqref{numberoperator} for the PDF) and for the decay constant.
For a spinless state, such as a meson, the electromagnetic form factor~$\FF(Q^2)$ is defined as~\cite{drellyan,BRODSKY1998299}:
\begin{equation}
    \label{FFdef}
    \begin{gathered}
    \sand{P^{\,\prime}}{J_\mu(0)}{P}
    =
    (P^{\,\prime}_\mu+P_{\mu})
    \FF(Q^2)
     \ ,\\
    q_\mu = P^{\,\prime}_\mu - P_\mu
     \ ,\qquad
    Q^2 = -q^\mu q_\mu
     \ .
    \end{gathered}
\end{equation}
Switching to the Drell-Yan frame implies directing the incident hadron along the $z$-axis, and setting photon's momentum $q^\mu$ transverse to this direction:
\begin{equation}
    \label{drellyan}
    P^\mu = (P^+,\,\vec{0}_\perp,\,M^2/P^+)
     \ ,\qquad
    q^\mu = (0,\,\vec{q}_\perp,\,2q\cdot P/P^+)
     \ ,
\end{equation}
where $M$ is the hadron's mass.

\begin{sloppypar}
In the LF the full Heisenberg current ${J_\mu(0)
}$ in eq.~\eqref{FFdef} can be set equal to the free quark current $j_\mu(0)$~\cite{BRODSKY1998299}. Similarly to~\eqref{FFdef}, one can define the electroweak form factor by replacing~$J_\mu(x)$ with the chiral current~${J^{\,5}_\mu(x) = \sum_\Modeflavor \charge_\Modeflavor \overline{\Psi}_\Modeflavor(x) \gamma_\mu \gamma^5 \Psi_\Modeflavor(x)}$.
In the LF, the expression for the form factor then takes the following form~\cite{frame}:
\end{sloppypar}
\begin{gather}
    \label{formfactordef}
    \FF(Q^2)
    =
    \dfrac{1}{\,2P^{\mathrlap+\;\;\,}}
    \sand{P^{\,\prime}}{J^+(0)}{P}
    =
    \dfrac{1}{\,2P^{\mathrlap+\;\;\,}}
    \sum_{\sket{\{\collective_j,\occ_j\}}}
    {\sum_{\!\!\mathrlap\collectivestruck}}^{\prime}
    \charge_{\collectivestruck} \sbraket{P^{}}{\{\collective_j',\occ_j\}_{\collectivestruck}}\sbraket{\{\collective_j,\occ_j\}}{P}
     \ ,
\end{gather}
where the first sum goes over all the Fock states, while $\collectivestruck$ indicates the choice of the struck quark in~$\sket{\{\collective_j,\occ_j\}}$ and varies over all the quark modes (with charges $\charge_{\collectivestruck}$). The state~$\sket{\{\collective_j',\occ_j\}_{\collectivestruck}}$ differs from~$\sket{\{\collective_j,\occ_j\}}$ only in its transverse momenta:
\begin{equation}
    \label{ffmomenta}
    \vec{l}_{\perp\,j} =
    \begin{cases}
    \vec{k}_{\perp\,j} - x_j \vec{q}_{\perp} + \vec{q}_{\perp}
     \  & \text{for the struck quark, ($\collective'_j=\collectivestruck$),}
    \\
    \vec{k}_{\perp\,j} - x_j \vec{q}_{\perp}
     \  & \text{for all other partons, ($\collective'_j\neq\collectivestruck$).}
    \end{cases}
\end{equation}

Note that the final expression for the form factor in~\eqref{formfactordef} involves state~$\sket{P}$, but not~$\sket{P^{\,\prime}}$~\cite{BRODSKY1998299,frame}. To describe the  corresponding measurement, we can formally rewrite~\eqref{formfactordef} as
\begin{equation}
    \label{ffmeasurementop}
    \begin{gathered}
    \FF(Q^2) =
    \dfrac{1}{\,2P^{\mathrlap+\;\;\,}} \sand{P}{\formfM(Q^2)}{P}
     \ ,\\
    \formfM(Q^2)
    = \!\!\!\!
    \sum_{\sket{\{\collective_j,\occ_j\}}}
    \!\!\!
    {\sum_{\!\!\mathrlap\collectivestruck}}^{\prime}
    \charge_{\collectivestruck}
    \saxe
    {\{\collective_j',\occ_j\}_{\collectivestruck}}
    {\{\collective_j,\occ_j\}}
     \ .
    \end{gathered}
\end{equation}

Note that~$\formfM(Q^2)$ is not a Hermitian operator, which should not surprise us since the form factor is generally allowed to take complex values. Nonetheless, the real and imaginary parts of~$\FF(Q^2)$ can be obtained by measuring the following Hermitian operators:
\begin{equation}
    \begin{alignedat}{99}
    &\Re \FF(Q^2) &&=
    && \dfrac{1}{\,2P^{\mathrlap+\;\;\,}}
    &&\sand{P}{\mathrlap{\,\dfrac{1}{2}}\hphantom{\dfrac{1}{2\iu}}\bigl[\formfM(Q^2)+\formfM^\dagger(Q^2)\bigr]}{P}
     \ ,\\
    &\Im \FF(Q^2) &&=
    && \dfrac{1}{\,2P^{\mathrlap+\;\;\,}}
    &&\sand{P}{\dfrac{1}{2\iu}\bigl[\formfM(Q^2)\tabularbin{-}\formfM^\dagger(Q^2)\bigr]}{P}
     \ .
    \end{alignedat}
\end{equation}

Such measurements can be performed efficiently (see the discussion in~App.~\ref{appimplementation} for the analogous case of PDFs). This would amount to constructing a circuit identifying the Fock state and transforming it according to~\eqref{ffmomenta}. Importantly, the final state of each mode in~\eqref{ffmomenta} depends only on its initial state, and is not conditioned on the rest of the modes. Moreover, since each initial Fock state is mapped onto a linear combination of a small number (at most~${\lfloor\sqrt{2K}\rfloor}$) of final states, the matrix~$\formfM(Q^2)$ is sparse.

The LF wave functions can also be used to calculate meson decay constants. For scalar~(s) and pseudoscalar~(p) mesons, those can be written in terms of the vector and axial quark current operators as~\cite{quarkonium,basislightmesons,BRODSKY1998299}:
\begin{equation}
    \label{decayconstdef}
    \begin{alignedat}{22}
    &\sand{0}{J_\mu(0)}{P_\text{s}}
    &&=
    &&P_{\mu} \, f_\text{s}
     \ &&,\qquad
    &&\sand{0}{J^{\,5}_\mu(0)}{P_\text{p}}
    &&=
    \iu P_{\mu} \, f_\text{p}
     \ &&.
    \end{alignedat}
\end{equation}
Since decay constants are linear in the wave function, their measurement has to be designed somewhat differently from that of PDFs and form factors, which are bilinear in the wave function (see eqs.~\eqref{pdfgen} and~\eqref{FFdef}).
Up to a constant coefficient dependent on a particular state, the decay constant is the integral of the wave function over all the two-particle Fock states. For example, for~$\pi^\pm$ one has~\cite{BRODSKY1998299}:
\begin{equation}
    \label{decayconstantwavefunction}
    f_{\pi} = 2\sqrt{3} \sum_{\!x,\,\vec{k}_{\mathrlap\perp}} \sbraket{\underbrace{\{x,\vec{k}_\perp\}}_{d},\underbrace{\{1-x,-\vec{k}_\perp\}}_{\bar{u}}}{P}  \ .
\end{equation}

Since only the magnitude of the decay constant is physically significant, its calculation reduces to evaluating $\bigl|\sbraket{v}{P}\bigr|$ for some fixed vector $\sket{v}$.
This measurement can be performed efficiently as long as one can efficiently prepare the state $\sket{v}$ from a computational basis state.
Since, as we noted above, the decay constant only requires integrating over two-particle Fock states, the vector $\sket{v}$ turns out to be itself a linear combination of two-particle Fock states.
Thus, it is indeed efficiently preparable.

\section*{Discussion and Perspectives\label{brightfuturesec}}
\addcontentsline{toc}{section}{\nameref{brightfuturesec}}

We have demonstrated several advantages of the light-front formulation for quantum simulation of quantum field theory. The qubit requirements in the light-front approach are greatly reduced as compared with those in equal-time quantization. This is due to the smaller number of physical degrees of freedom, and the fact that the sum of occupancies in a Fock state is upper bounded by $K$ for fixed harmonic resolution $K$.

The Hamiltonian matrix at fixed harmonic resolution in the LF formalism is sparse~\cite{VaryAbInitio,BRODSKY1998299}, enabling us to make use of optimal simulation algorithms~\cite{berry2017exponential,berry2019time}. These algorithms require $\widetilde{O}(tK^4)$ gates to simulate time evolution for time $t$, with logarithmic dependence on error. For state preparation by simulation of adiabatic evolution, in the case of adiabatic paths whose max norm is bounded by $K$, we require $\widetilde{O}(TK^4)$ gates. Proving that such paths in fact obey the adiabatic theorem is a topic for future work.

The LF formalism allows one to calculate various measurable quantities directly from the bound state wave functions. We demonstrated how such observables can be efficiently calculated on a quantum computer, using as our main example the analogue of the parton distribution function. Quantum computation of these observables has been considered by other authors prior to this work in~\cite{mueller2019deeply,Lamm:2019uyc}. Some of the advantages of the light-front discussed in detail here were presented in~\cite{freese}. We hope future work will further develop all approaches to these problems.

In $(1+1)D$ for harmonic resolution $K$ the qubit requirements scale as $O(\sqrt{K}\log K)$ in the compact encoding, which is optimal up to logarithmic factors. The compact encoding of light-front Fock states was shown to be extendable to higher-dimensional field theories. The qubit scaling increases to $\widetilde{O}(K)$, which is a significant improvement compared to equal-time quantization. For a $20^3$ grid in momentum space with~${n_f=5}$ and~${n_c=3}$, equation~\eqref{qcdestimate} gives~$1360$ qubits~--- much less than ${4\times10^5}$ qubits on the grid of the same size in equal time quantization estimated in~\cite{Lamm:2019uyc}. This is comparable to the number of logical qubits required to factor a $1024$ bit RSA key using Shor's algorithm~\cite{gidney2019factor}.

In higher dimensions, more observables can be calculated within a Hamiltonian block of a fixed longitudinal momentum. Those include decay constants, form-factors, generalized parton distributions functions, transverse-momentum-dependent distributions. As a possible direction of future work one could consider direct evaluation of the hadronic tensor. Instead of calculating it perturbatively (with the zeroth order being the parton-model approximation considered in the present paper), one could switch to the interaction picture, thus keeping the measurement operators unchanged while slightly complicating the state preparation.

Further development and optimization of the simulation techniques is warranted. Encoding schemes that restrict to a particular block of both $K$ and $Q$ would not change the asymptotic scaling of the qubit requirements but might be practically useful. Similarly, improvements to the implementation schemes given herein could yield significant reduction in gate numbers even if scaling improvements cannot be achieved. Such improvements likely require the development of software allowing the simulation of this algorithm, as has been developed for quantum algorithms for quantum chemistry~\cite{mcclean2017openfermion}.

An important issue arising within the DLCQ approach to QCD, which we have not addressed in the present work, is the effect of gluon zero-modes, whose absence in the free-field expansion is critical to our ability to use the compact encoding scheme. Zero-modes play an important role in the light-front formulation, incorporating all the complexity of the theory related to the non-triviality of the vacuum in the equal-time formulation~\cite{Kalloniatis:1993jh,Kalloniatis:1994ei,Kalloniatis:1994fk,Kalloniatis:1994ix,Kalloniatis:1995fc,Pauli:1995dt}. In the context of DLCQ, taking zero-modes into account may result in the appearance of new, non-canonical interactions~\cite{BRODSKY1998299}. As noted in~\cite{BRODSKY1998299}, while the longitudinal confinement is immanent in the light-cone quanitization of QCD due to the linear growth of effective potential in the $x^-$ direction, the interactions arising from zero-modes may be responsible for the transverse confinement.

As an alternative to the fundamental QCD Lagrangian, one can use effective low-energy theories. At the level of simulation, this amounts to changing the set of basis states to the one better resembling the bound state wave function. The latter approach seems to be particularly appealing due to the recent success of the so-called basis light-front quantization (BLFQ) technique~\cite{varybasis}. Within this method, the effective Lagrangian, respecting all the symmetries of the full QCD, is solved in the basis provided by an exactly solvable model emerging from the \mbox{AdS/QCD}~\cite{Brodsky:2008gc,Brodsky:2010ev,Brodsky:2012hy,BRODSKY1998299,karchkatz,holography,VEGA1,VEGA2,VEGA3,VEGA4,VEGA5,BRAGA2016203,BrodskyLightFrontDynamics}. We leave these, and other further details of the application to $3+1D$, to future work.


\section*{Acknowledgements}
M.~K., P.~J.~L. and G.~G. acknowledge support from DOE HEP Grant No. DE-SC0019452.
W.~M.~K. acknowledges support from NSF GRFP Grant No. DGE-1842474.
This work was supported by the NSF STAQ project (PHY-1818914). The authors wish to thank  Prof. Stanley Brodsky and Prof. James Vary for helpful discussions.

\begin{appendices}


\section{Hamiltonian\label{appham}}
Following~\cite{pauli1,pauli2}, we write the Hamiltonian as
\begin{equation}
    H = H_M + H_V + H_S + H_F  \ ,
\end{equation}
where
{\savebox\strutbox{$\vphantom{\sum\limits_{k}}$}
\begin{alignat}{99}
    \label{hmass}
    &H_M = \sum_\Moden \dfrac{1}{\Moden} \left[
    \OPad_\Moden \OPa_\Moden (\mass_B^2 + g^2 \alpha_\Moden)
    + \OPbd_\Moden \OPb_\Moden (\mass_F^2 + g^2 \beta_\Moden)
    + \OPdd_\Moden \OPd_\Moden (\mass_F^2 + g^2 \gamma_\Moden) \right]
     \ ,
\\
    \label{hvertex}&
    \begin{alignedat}{99}
    H_V = g \mass_F \sum_{\Modek,\Model,\Modem} \Bigl[\Bigr.
    &(\OPbd_\Modek \OPb_\Modem
    \OPcd_\Model + \OPbd_\Modem \OPb_\Modek \OPc_\Model)
    &&\left( \{\Modek+\Model|-\Modem\}+\{\Modek|+\Model-\Modem\} \right) \\
    +
    &(\OPdd_\Modek \OPd_\Modem \OPcd_\Model + \OPdd_\Modem \OPd_\Modek \OPc_\Model)
    &&\left( \{\Modek+\Model|-\Modem\}+\{\Modek|+\Model-\Modem\} \right) \\
    +
    &(\OPb_\Modek \OPd_\Modem \OPcd_\Model + \OPdd_\Modem \OPbd_\Modek \OPc_\Model)
    &&\left( \{\Modek-\Model|+\Modem\}+\{\Modek|-\Model+\Modem\} \right) \Bigl.\Bigr] \ ,
    \end{alignedat}
\\
    \label{hseagull}
    &
    \begin{alignedat}{99}
    H_S = g^2 \sum_{\Modek,\Model,\Modem,\Moden} \Bigl[\Bigr.
    &\OPbd_\Modek \OPb_\Modem \OPcd_\Model \OPc_\Moden
    \left( \{\Modek-\Moden|\Model-\Modem\}+\{\Modek+\Model|-\Modem-\Moden\} \right) \\
    +
    &\OPdd_\Modek \OPd_\Modem \OPcd_\Model \OPc_\Moden
    \left( \{\Modek-\Moden|\Model-\Modem\}+\{\Modek+\Model|-\Modem-\Moden\} \right) \\
    +
    &(\OPd_\Modek \OPb_\Modem \OPcd_\Model \OPcd_\Moden +\OPbd_\Modem \OPdd_\Modek \OPc_\Moden \OPc_\Model)
    \{\Model-\Modek|\Moden-\Modem\}
    \Bigl.\Bigr] \ ,
    \end{alignedat}
\\
    \label{hfork}
    &
    \begin{alignedat}{99}
    H_F = g^2 \sum_{\Modek,\Model,\Modem,\Moden} &\Bigl[\Bigr.
    &&(&&\OPbd_\Modek \OPb_\Modem \OPcd_\Model \OPcd_\Moden &&+\OPbd_\Modem \OPb_\Modek \OPc_\Moden \OPc_\Model)
    \{\Modek+\Model|\Moden-\Modem\} \\
    &+
    &&(&&d_\Modek^\sdagger \OPd_\Modem \OPcd_\Model \OPcd_\Moden &&+d_\Modem^\sdagger \OPd_\Modek \OPc_\Moden \OPc_\Model)
    \{\Modek+\Model|\Moden-\Modem\} \\
    &+
    &&&&\OPbd_\Modek \OPdd_\Modem \OPcd_\Model \OPc_\Moden \bigl( &&\{\Modek-\Moden|\Modem+\Model\} + \{\Modek+\Model|\Modem-\Moden\}\bigr) \\
    &+
    &&&&\OPd_\Modem \OPb_\Modek \OPcd_\Moden \OPc_\Model \bigl( &&\{\Modek-\Moden|\Modem+\Model\} + \{\Modek+\Model|\Modem-\Moden\}\bigr)
    \Bigl.\Bigr] \ ,
    \end{alignedat}
\end{alignat}
}
\begin{sloppypar*}
where ${\OPc_\Moden = \OPa_\Moden / \sqrt{\Moden}}$. The expressions in \mbox{\eqref{hmass}-\eqref{hfork}} are called the mass, vertex, seagull and fork parts of the Hamiltonian, respectively.
\end{sloppypar*}

\subsection{Self-induced inertias}

The mass term contains the so-called self-induced inertias~$\alpha_\Moden,\,\beta_\Moden,\,\gamma_\Moden$, the cutoff-dependent quantities whose appearance is a general phenomenon in the LF framework not specific to the particular theory under consideration. Those are defined as
\begin{equation}
\begin{gathered}
    \label{intertias}
    \alpha_\Moden = \sum_{\Modem=1}^\Lambda \left(
    \{\Moden-\Modem|\Modem-\Moden\} - \{\Moden+\Modem|-\Modem-\Moden\}
    \right)  \ ,\\
    \beta_\Moden = \sum_{\Modem=1}^\Lambda \dfrac{\Moden}{\Modem} \{\Moden-\Modem|\Modem-\Moden\}  \ , \qquad
    \gamma_\Moden = \sum_{\Modem=1}^\Lambda \dfrac{\Moden}{\Modem} \{\Moden+\Modem|-\Modem-\Moden\}  \ .
\end{gathered}
\end{equation}
where
\begin{equation}
    \label{bracket}
    \{\Moden | \Modem\} = \begin{cases}
    0  \ &\text{if } \Moden=0 \text{ or } \Modem=0  , \\
    \dfrac{1}{\Moden} \delta_{\Modem,-\Moden}  \ & \text{otherwise}.
    \end{cases}
\end{equation}
\begin{sloppypar}
We must upper bound these quantities as they contribute to the norm of the Hamiltonian, which in turn determines the simulation complexity. We can first evaluate the sums as follows:
\end{sloppypar}
\begin{equation}
\begin{alignedat}{9}
\label{harmonicintertias}
\alpha_\Moden &=  -\frac{1}{n} - H_{\Lambda - \Moden} - H_{2\Moden} +
 2 H_{\Moden} \ &&,\\
\beta_\Moden &= -\frac{2}{n} + H_{\Moden} + H_\Lambda - H_{\Lambda - \Moden} \ &&,\\
\gamma_\Moden &= -\frac{1}{2\Moden} + H_\Moden + H_{\Lambda}-H_{\Lambda+n} \ &&,\\
\end{alignedat}
\end{equation}
where $H_n$ is the $n$th harmonic number. Using the well-known bounds on the harmonic numbers we can upper and lower bound the self induced inertias as a function of the cutoff~\cite{pauli2}:
\begin{equation}
\label{boundintertias}
\begin{alignedat}{9}
\log\frac{\Moden}{2 (\Lambda - \Moden)}-2 + \frac{1}{\Moden} &\leq\alpha_\Moden &&\leq \log\frac{n}{2 (\Lambda - n)}+2 - \frac{3}{2n} - \frac{1}{(\Lambda - n)} \ &&,\\
\log\frac{\Lambda \Moden}{\Lambda - \Moden}-1-\frac{1}{\Moden}+\frac{1}{\Lambda}&\leq\beta_\Moden &&\leq  \log\frac{\Lambda \Moden}{\Lambda - \Moden} + 2 - \frac{2}{\Moden} - \frac{1}{\Lambda - \Moden} \ &&,\\
\log\frac{\Moden \Lambda}{\Lambda +\Moden} -1 + \frac{1}{2\Moden} + \frac{1}{\Lambda + \Moden} &\leq\gamma_\Moden   &&\leq \log\frac{\Moden \Lambda}{\Lambda +\Moden}+2 - \frac{1}{2\Moden} - \frac{1}{\Lambda + \Moden} \ &&.\\
\end{alignedat}
\end{equation}
Because $n\leq K$ thes bounds show that all self-induced inertias scale with $K$ and $\Lambda$ as~$\Theta(\log(K/\Lambda))$.


\subsection{Hamiltonian sparsity\label{appsparse}}

The \emph{sparsity} of the Hamiltonian in the Fock basis is the maximum number of final states onto which a single initial Fock state is mapped under the action of the Hamiltonian. The mass terms in the Hamiltonian, $H_M$ \eqref{hmass}, are proportional to number operators, hence are diagonal in the Fock basis, and so do not contribute to the sparsity. We analyze the sparsity of the terms in the Hamiltonian, $H_V$ \eqref{hvertex}, $H_S$ \eqref{hseagull}, and $H_F$ \eqref{hfork}, by separately finding the numbers of nonzero images of each set of terms with a given form, for a generic initial state, then summing these and maximizing the resulting expression. We call the state whose number of nonzero images is maximal the \emph{sparsity-determining state} (SDS).

Each Hamiltonian term contains annihilation operators that will map a state to zero unless they act on an occupied mode.
Thus the number of nonzero images is largest for a state in which all bosons have distinct momenta, because this maximizes the number of Hamiltonian terms in which the annihilation operators do not map the state to zero.
Therefore, all occupation numbers in the SDS are zero or one, so the SDS is determined by the sets $\Forb$, $\Aorb$, and $\Borb$ of occupied fermionic, antifermionic, and bosonic momenta (respectively).

To obtain an upper bound on the sparsity, we assume that every term whose annihilation operators act on occupied modes maps the initial state to a distinct nonzero image. This is a relaxation of the actual condition in two respects: first, some of the nonzero images thus obtained may not be distinct, and second, fermionic and antifermionic creation operators acting on \emph{occupied} modes will map the state to 0 rather than to a nonzero image. However, we will see that the upper bound we obtain by ignoring these reductions to the sparsity will nonetheless turn out to be asymptotically tight.

We consider sets of terms of a fixed form, e.g., $\{\OPbd_k\OPb_m\OPcd_l~|~k,l,m\in\{1,2,...,K\},~k+l=m\}$.
These sets are represented in eqs.~\eqref{spar_v_1}-\eqref{spar_f_6} by a characteristic element, e.g., $\OPbd_k\OPb_m\OPcd_l$.
In each set, the modes for each ladder operator vary over ${\{1,2,...,K\}}$, under the constraint that total momentum is conserved, i.e., the sum of the momenta of the annihilation operators is equal to the sum of the momenta of the creation operators, which is equal to the transferred momentum. Each term is thus associated to a particular value of transferred momentum.

For each set of terms, the transferred momentum will be a sum over the possible sets of occupied modes corresponding to the annihilation operators in the set of terms. The summand will be the sum over the possible assignments of the transferred momentum to the creation operators in the set of terms.
We can thus tabulate the numbers of nonzero images for each set of terms.
We will use the following facts repeatedly: if some transferred momentum $m$ is to be divided between two outgoing modes, this may be accomplished in $m-1$ ways (since each mode must have nonzero momentum), obtaining $m-1$ nonzero images.
If there is only one outgoing mode, then clearly it must possess all of the transferred momentum, giving only one nonzero image.
The numbers of nonzero images for each set of terms are given by:
\begin{subequations}
\begin{alignat}{9}
    &\OPbd_k\OPb_m\OPcd_l\quad&&\Rightarrow&&\quad\sum_{m\in\Forb}(m-1)=\sum_{m\in\Forb}(m)-|\Forb| \ ,\label{spar_v_1}\\
    &\OPdd_k\OPd_m\OPcd_l\quad&&\Rightarrow&&\quad\sum_{m\in\Aorb}(m-1)=\sum_{m\in\Aorb}(m)-|\Aorb| \ ,\label{spar_v_2}\\
    &\OPbd_k\OPdd_m\OPc_l\quad&&\Rightarrow&&\quad\sum_{l\in\Borb}(l-1)=\sum_{m\in\Borb}(m)-|\Borb| \ ,\label{spar_v_3}\\
    &\OPb_k\OPbd_m\OPc_l\quad&&\Rightarrow&&\quad\sum_{k\in\Forb}\sum_{l\in\Borb}1=|\Forb||\Borb| \ ,\label{spar_v_4}\\
    &\OPd_k\OPdd_m\OPc_l\quad&&\Rightarrow&&\quad\sum_{k\in\Aorb}\sum_{l\in\Borb}1=|\Aorb||\Borb| \ ,\label{spar_v_5}\\
    &\OPb_k\OPd_m\OPcd_l\quad&&\Rightarrow&&\quad\sum_{k\in\Forb}\sum_{m\in\Aorb}1=|\Forb||\Aorb| \ ,\label{spar_v_6}\\
    &\OPbd_k\OPb_m\OPcd_l\OPc_n\quad&&\Rightarrow&&\quad\sum_{m\in\Forb}\sum_{n\in\Borb}(m+n-1)=|\Borb|\sum_{m\in\Forb}(m)+|\Forb|\sum_{n\in\Borb}(n)-|\Forb||\Borb| \ ,\label{spar_s_1}\\
    &\OPdd_k\OPd_m\OPcd_l\OPc_n\quad&&\Rightarrow&&\quad\sum_{m\in\Aorb}\sum_{n\in\Borb}(m+n-1)=|\Borb|\sum_{m\in\Aorb}(m)+|\Aorb|\sum_{n\in\Borb}(n)-|\Aorb||\Borb| \ ,\label{spar_s_2}\\
    &\OPbd_m\OPdd_k\OPc_n\OPc_l\quad&&\Rightarrow&&\quad\sum_{l\in\Borb}\sum_{n\in\Borb}(l+n-1)=2|\Borb|\sum_{n\in\Borb}(n)-|\Borb|^2 \ ,\label{spar_s_3}\\
    &\OPb_m\OPd_k\OPcd_l\OPcd_n\quad&&\Rightarrow&&\quad\sum_{m\in\Forb}\sum_{k\in\Aorb}(m+k-1)=|\Aorb|\sum_{m\in\Forb}(m)+|\Forb|\sum_{n\in\Aorb}(n)-|\Forb||\Aorb| \ ,\label{spar_s_4}\\
    &\OPbd_k\OPb_m\OPcd_l\OPcd_n\quad&&\Rightarrow&&\quad\sum_{m\in\Forb}\sum_{k=1}^{m-2}(m-k-1)=\sum_{m\in\Forb}\frac{(m-1)(m-2)}{2}=\sum_{m\in\Forb}\left(\frac{m^2-3m}{2}\right)+|\Forb| \ ,\label{spar_f_1}\\
    &\OPdd_k\OPd_m\OPcd_l\OPcd_n\quad&&\Rightarrow&&\quad\sum_{m\in\Aorb}\sum_{k=1}^{m-2}(m-k-1)=\sum_{m\in\Aorb}\frac{(m-1)(m-2)}{2}=\sum_{m\in\Aorb}\left(\frac{m^2-3m}{2}\right)+|\Aorb| \ ,\label{spar_f_2}\\
    &\OPbd_k\OPdd_n\OPcd_l\OPc_m\quad&&\Rightarrow&&\quad\sum_{m\in\Borb}\sum_{k=1}^{m-2}(m-k-1)=\sum_{m\in\Borb}\frac{(m-1)(m-2)}{2}=\sum_{m\in\Borb}\left(\frac{m^2-3m}{2}\right)+|\Borb| \ ,\label{spar_f_3}\\
    &\OPbd_m\OPb_k\OPc_n\OPc_l\quad&&\Rightarrow&&\quad\sum_{k\in\Forb}\sum_{n\in\Borb}\sum_{l\in\Borb}1=|\Forb||\Borb|^2 \ ,\label{spar_f_4}\\
    &\OPbd_m\OPb_k\OPc_n\OPc_l\quad&&\Rightarrow&&\quad\sum_{k\in\Aorb}\sum_{n\in\Borb}\sum_{l\in\Borb}1=|\Aorb||\Borb|^2 \ ,\label{spar_f_5}\\
    &\OPd_m\OPb_k\OPcd_n\OPc_l\quad&&\Rightarrow&&\quad\sum_{k\in\Forb}\sum_{m\in\Aorb}\sum_{l\in\Borb}1=|\Forb||\Aorb||\Borb| \ .\label{spar_f_6}
\end{alignat}
\end{subequations}
Here the term on the left is the representative element of an entire set of terms of that type.

Our upper bound on the total number of nonzero images of the full Hamiltonian is the sum of these:
\begin{equation}
\label{sparsityinit}
\begin{multlined}
    \sum_{m\in\Forb}(m)+\sum_{m\in\Aorb}(m)+\sum_{m\in\Borb}(m)\\
    +\sum_{m\in\Forb}\sum_{n\in\Borb}(m+n)+\sum_{m\in\Aorb}\sum_{n\in\Borb}(m+n)+\sum_{m\in\Borb}\sum_{n\in\Borb}(m+n)+\sum_{m\in\Forb}\sum_{n\in\Aorb}(m+n)\\
    +\sum_{m\in\Forb}\left(\frac{m^2-3m}{2}\right)+\sum_{m\in\Aorb}\left(\frac{m^2-3m}{2}\right)+\sum_{m\in\Borb}\left(\frac{m^2-3m}{2}\right)\\
    +|\Forb||\Borb|^2+|\Aorb||\Borb|^2+|\Forb||\Aorb||\Borb|
    -|\Borb|^2\mathrlap{ \ .}
\end{multlined}
\end{equation}

To simplify the above expression, let
\begin{equation}
    K_F\equiv\sum_{m\in\Forb}m \ ,\qquad
    K_A\equiv\sum_{m\in\Aorb}m \ ,\qquad
    K_B\equiv\sum_{m\in\Borb}m \ ,
\end{equation}
\begin{sloppypar*}
denote the total momenta possessed by fermions, antifermions, and bosons (respectively) in the initial state.
The sum of these must be the total momentum, i.e., ${K_F+K_A+K_B=K}$.
Furthermore, by \eqref{maxnmodes}, the constraints on the sizes of the sets of momenta are ${1\le|\Forb|=I_F\le\sqrt{2K_F}}$, ${1\le|\Aorb|\le\sqrt{2K_A}}$, and ${1\le|\Borb|\le\sqrt{2K_B}}$.
\end{sloppypar*}

Thus our upper bound on the number of nonzero images of the Hamiltonian becomes:
\begin{equation}
\label{sparsity2}
\begin{multlined}
    \frac{1}{2}\sum_{m\in\Forb}(m^2)+\frac{1}{2}\sum_{m\in\Aorb}(m^2)+\frac{1}{2}\sum_{m\in\Borb}(m^2)+(|\Forb|+|\Aorb|-1)|\Borb|^2+|\Forb||\Aorb||\Borb|\\
    +K_F|\Borb|+K_B|\Forb|+K_A|\Borb|+K_B|\Aorb|+2K_B|\Borb|+K_F|\Aorb|+K_A|\Forb|-\frac{1}{2}K\\
    =\frac{1}{2}\sum_{m\in\Forb}(m^2)+\frac{1}{2}\sum_{m\in\Aorb}(m^2)+\frac{1}{2}\sum_{m\in\Borb}(m^2)+(|\Forb|+|\Aorb|-1)|\Borb|^2+|\Forb||\Aorb||\Borb|\\
    \quad+(K-K_F)|\Forb|+(K-K_A)|\Aorb|+(K+K_B)|\Borb|-\frac{1}{2}K\mathrlap{ \ .}
\end{multlined}
\end{equation}

Since $|\Forb|,|\Aorb|,|\Borb|$ scale at most as the square root of $K$, only the first three terms in this expression grow as $K^2$: all others grow at most as $K^{3/2}$. Thus for large $K$, the sparsity is maximized by maximizing the first three terms in \eqref{sparsity2}:
\begin{equation}
\label{firstthreeterms}
    \frac{1}{2}\sum_{m\in\Forb}(m^2)+\frac{1}{2}\sum_{m\in\Aorb}(m^2)+\frac{1}{2}\sum_{m\in\Borb}(m^2) \ .
\end{equation}
But this expression is clearly maximized when all of the initial momentum is carried by a single particle, i.e., one of $\Forb$, $\Aorb$, or $\Borb$ is $\{K\}$, and the other two are empty. Which we should choose is determined by the remaining terms in \eqref{sparsity2}: the maximizing choice is ${\Borb=\{K\}}$, ${|\Forb|=|\Aorb|=0}$. Substituting these assignments into \eqref{sparsity2} gives
\begin{equation}
\label{sparsityupper}
    \frac{1}{2}K^2+\frac{3}{2}K-1 \ ,
\end{equation}
which is thus our upper bound for the sparsity. Direct evaluation of the sparsity of the Hamiltonian for small $K$ shows that this bound holds for all $K$. The results are plotted in Fig.~\ref{sparsityfig}.

To obtain a lower bound, note that out of all contributions to the sparsity in \eqref{spar_v_1}-\eqref{spar_f_6}, the largest is
\begin{equation}
    \sum_{m\in\Borb}\left(\frac{m^2-3m}{2}\right)+|\Borb| \ ,
\end{equation}
in \eqref{spar_f_3}; the maximizing term type is $\OPbd_k\OPdd_n\OPcd_l\OPc_m$.
We choose this set of terms rather than that in \eqref{spar_f_1} or \eqref{spar_f_2}, because for the terms $\OPbd_k\OPdd_n\OPcd_l\OPc_m$ all creation operators act on different kinds of particles, which removes the possibility of double-counting nonzero images.
The SDS that maximizes \eqref{spar_f_3} is the same as the SDS for the full Hamiltonian: a single initial boson with momentum $K$.
Therefore, the fermion and antifermion creation operators $\OPbd_k,\OPdd_n$ can create a nonzero image for each allowed value of $(k,n)$, since all initial fermion and antifermion occupation numbers are zero.
Thus, the sparsity of the set of terms ${\OPbd_k\OPdd_n\OPcd_l\OPc_m}$ is \emph{exactly} the value of \eqref{spar_f_3} when ${\Borb=\{K\}}$, and forms our lower bound on the sparsity of the full Hamiltonian:
\begin{equation}
    \label{sparsitylower}
    \spam=\sum_{m\in\Borb}\left(\frac{m^2-3m}{2}\right)+|\Borb|=\frac{1}{2}K^2-\frac{3}{2}K+1 \ .
\end{equation}

Fig.~\ref{sparsityfig} (in Sec.~\ref{statesandops}) plots the upper and lower bounds together with exact values for the sparsity, which we calculated directly from the Hamiltonians with ${Q=0}$ and ${K=3,4,\ldots,19}$. The sparsity of the Hamiltonian is therefore tightly bounded by eq.~\eqref{sparsityupper} and eq.~\eqref{sparsitylower}, is ${\frac{1}{2}K^2}$ at leading order in $K$, and is $\Theta(K^2)$.


\subsection{Hamiltonian norm\label{appnorm}}

We define $\snormmax{H}$ as the largest matrix element of $H$ in absolute value:
\begin{equation}
    \label{normdef}
    \snormmax{H} = \underset{j,k}{\max}|H_{j,k}| \ .
\end{equation}
As follows from eq.~\eqref{hmass}, the Fock states having largest eigenvalues (considered as the eigenstates of the free Hamiltonian) are the ones with high bosonic occupancy, such as~${\sket{;;\widetilde{1}^{K-2},\widetilde{2}}}$ or ~${\sket{1;\overline{1};\widetilde{1}^{K-2}}}$\footnote{
We do not use~$\sket{;;\widetilde{1}^K}$ for it is the so-called \emph{angel state} which decouples from the rest of the spectrum~\cite{pauli2}.}. When acting on those, the number operator~$\OPad_\Moden \OPa_\Moden$ produces a factor of order~$O(K)$.

The same logic is applicable to the interaction terms~\eqref{hvertex}-\eqref{hfork}, none of which can result in more than a linear dependence on $K$. Lastly, the self-induced inertias scale asymptotically with $K$ and $\Lambda$ as~$O(\log K/\Lambda)$~\eqref{boundintertias}. Altogether, this results in the following bound for the Hamiltonian norm:
\begin{equation}
    \label{hamiltoniannormmax}
    \snormmax{H} =  O(K \log K/\Lambda)= \widetilde{O} (K) \ .
\end{equation}
We may use the relations amongst the various Hamiltonian norms in~\cite{childs2010limitations} and the sparsity of the Hamiltonian from~\ref{appsparse} to bound the spectral norm of $H$ as:
\begin{equation}
    \label{hamiltoniannorm}
    ||H|| =  O(K^3 \log K/\Lambda)= \widetilde{O} (K^3) \ .
\end{equation}


\section{Compact Mapping}
\label{partitions}

In this appendix, we describe how to efficiently encode an occupation number state (i.e., a partition of momentum) in $1+1D$ at fixed Harmonic resolution as a qubit state.
This construction may be applied to bosons (with occupancies in $[0,K]$), or fermions, or antifermions (with occupancies in $\{0,1\}$).
We encode an occupation number state as a list of pairs: ${\{(n_1,\occ_1),(n_2,\occ_2),...,(n_I,\occ_I)\}}$,
where the $n_i$ are the \emph{distinct} momenta (part sizes) that appear, and the ${\occ_i=[1,K]}$ are the corresponding occupancies numbers (we do not store modes with $w_i=0$).
Each $n_i,\occ_i$ is an integer in $[1,K]$, so we encode each pair $(n_i,m_i)$ in a register of size ${2\lceil\log_2K\rceil}$ qubits.
The total number of qubits required is then

\begin{equation}
    \label{partitionqubits}
    \qubitn_K=2I\lceil\log_2K\rceil \ ,
\end{equation}
since we use $I$ registers.

Label the $i$th register $X_i$: then the complete qubit states used to encode the momentum partitions may be written ${|X_1,X_2,...,X_I\rangle}$.
$I$ should be equal to the maximum number of distinct part lengths in any partition of $K$, but most partitions will not have $I$ distinct part lengths, so in general the states will not use all of the registers $X_I$.
In order to uniquely associate an encoding state to each partition, we choose the following conventions:
\begin{enumerate}
    \item Occupied momenta will be arranged in decreasing order of size.
    \item If $I'\le I$ momenta are occupied, they are encoded in the first $I'$ of the $X_i$.
\end{enumerate}

How do we determine $I$, the maximum number of distinct part lengths in any partition of $K$?
Simply adding up the momenta gives
\be
\sum_{i=1}^I n_i \occ_i=K \ ,
\ee
so we may obtain a tight bound on $I$ by noting that the least value of $K$ for a given $I$ is obtained by setting ${n_i=i}$ and ${\occ_i=1}$ for ${i\in\{1,2,...,I\}}$.
In this case, $K$ is a triangular number, i.e.,
\be
\label{triangularcase}
K=\sum_{i=1}^I n_i \occ_i=\sum_{i=1}^I i=\frac{I(I+1)}{2} \ ,
\ee
so
\be
\label{maxnmodes}
I\le\sqrt{2K} \ .
\ee

The bound \eqref{maxnmodes} is satisfactory for analyzing the asymptotic qubit requirements, but to set the number of qubits we want to choose the minimum possible integer $I$ such that a partition of $K$ contains at most $I$ distinct parts: this turns out to be
\begin{equation}
    \label{minI}
    I=\left\lfloor\sqrt{2K+\frac{1}{4}}-\frac{1}{2}\right\rfloor \ .
\end{equation}
To see that \eqref{minI} gives the minimal $I$, let $K^\triangle$ be the largest triangular number less than or equal to $K$; then the minimal $I$ exactly satisfies
\begin{equation}
    \frac{I(I+1)}{2}=K^\triangle \ ,
\end{equation}
since for $K^\triangle$, \eqref{triangularcase} applies exactly.
From this we obtain
\begin{equation}
    4I^2+4I=8K^\triangle \quad \Rightarrow \quad
    (2I+1)^2=8K^\triangle+1 \ ,
\end{equation}
i.e., ${8K^\triangle+1}$ is an odd square.
This implication reverses: if ${8K'+1=J^2}$ for odd $J$, then we can choose ${I=\frac{1}{2}(J-1)}$ and we obtain ${K'=\frac{I(I+1)}{2}}$, so $K'$ is triangular.
The largest odd integer less than or equal to some arbitrary $x$ is ${2\lfloor\frac{x-1}{2}\rfloor+1}$, so for arbitrary $K$ the largest odd $J$ whose square is less than or equal to ${8K+1}$ is:
\begin{equation}
    J=2\left\lfloor\frac{\sqrt{8K+1}-1}{2}\right\rfloor+1 \ .
\end{equation}
Thus ${8K^\triangle+1=J^2}$ for $J$ determined in this way, so
\begin{equation}
\begin{gathered}
    2I+1=J=2\left\lfloor\frac{\sqrt{8K+1}-1}{2}\right\rfloor+1
    \\\mathllap{\Rightarrow} \quad
    I=\left\lfloor\sqrt{2K+1/4}-\frac{1}{2}\right\rfloor\mathrlap{ \ ,}
\end{gathered}
\end{equation}
which is \eqref{minI}.

For example, suppose the momentum is $K=6$ (chosen to be a triangular number, for convenience).
Then $I=3$, and the possible partitions are encoded as
\begin{equation}
\begin{alignedat}{8}
    |X_1,X_2,X_3\rangle=~&|(6,1),(0,0),(0,0)\rangle \ &&,\\
    &|(5,1),(1,1),(0,0)\rangle \ &&,\\
    &|(4,1),(2,1),(0,0)\rangle \ &&,\\
    &|(4,1),(1,2),(0,0)\rangle \ &&,\\
    &|(3,2),(0,0),(0,0)\rangle \ &&,\\
    &|(3,1),(2,1),(1,1)\rangle \ &&,\\
    &|(3,1),(1,3),(0,0)\rangle \ &&,\\
    &|(2,3),(0,0),(0,0)\rangle \ &&,\\
    &|(2,2),(1,2),(0,0)\rangle \ &&,\\
    &|(2,1),(1,4),(0,0)\rangle \ &&,\\
    &|(1,6),(0,0),(0,0)\rangle \ &&,
\end{alignedat}
\end{equation}
where each ${X_i=(n_i,\occ_i)}$ is encoded in ${2\lceil\log_2(6)\rceil=6}$ qubits, 3 to encode each of $n_i$ and $\occ_i$.

In the case of our algorithm the momentum is partitioned among fermions, antifermions, and bosons.
Let $K$ continue to denote the total momentum, summed over the fermions, antifermions, and bosons.
For bosons, we use exactly the mapping of Fock states to qubit states described above; we still require ${I_{\text{bosons}}=I}$ as given in \eqref{minI}, since in some states all of the momentum $K$ is possessed by bosons.
For fermions and antifermions, we use the mapping described above, but with the occupation numbers restricted to be 0 or 1.
Since only momenta that are present are represented in the state, this means that for all momenta that are present, $\occ_i$ restricted to be 1.
Thus we may drop the occupation numbers $\occ_i$ entirely, and simply keep a list of the fermion and antifermion momenta that are present.
We still require ${I_{\text{fermions}}=I_{\text{antifermions}}=I}$ as given in \eqref{minI}, since in some states all of the momentum $K$ is possessed by fermions or antifermions.
Thus our complete Fock states are stored as
\begin{equation}
    \{n_1,n_2,...,n_I;~\overline{n}_1,\overline{n}_2,...,\overline{n}_I;~(\widetilde{n}_1,\widetilde{\occ}_1),(\widetilde{n}_2,\widetilde{\occ}_2),...,(\widetilde{n}_I,\widetilde{\occ}_I)\} \ ,
\end{equation}
where $n_i,\overline{n}_i,\widetilde{n}_i\in\{1,...,K\}$ denote the fermion, antifermion, and boson momenta that are present in the state, and $\widetilde{\occ}_i\in\{1,...,K\}$ denote the occupation numbers of the occupied boson momenta.
Thus the total number of qubits that this encoding requires is
\begin{equation}
    \label{singlestatetotalqubits}
    \qubitn=4I\lceil\log_2K\rceil\le4\sqrt{2K}\lceil\log_2K\rceil
    \ .
\end{equation}


\section{Implementation\label{appimplementation}}

In this section we describe the details of the implementation of two oracles necessary for the sparse simulation algorithm in $1+1D$. These oracles are ubiquitous in methods for simulation of sparse Hamiltonians and were defined for the elctronic structure problem in~\cite{babbush2016exponentially,Toloui}. There are two differences in the definition of these oracles for the model defined in Sec.~\ref{themodelsection}. Firstly, we do not rely on any analogue of the Slater rules that define the nonzero matrix elements of the configuration-interaction (CI) matrix. Instead we use the second quantized representation of the Hamiltonian to enumerate nonzero elements of a row or column. Secondly, for the electronic structure Hamiltonian the matrix elements are defined in terms of integrals over basis functions whereas ours are simple functions of the momenta.

One could adapt the methods of~\cite{babbush2016exponentially,Toloui} to the enumeration of nonzero matrix elements. This would required the computation of the analog of the Slater rules and their implementation in the Hamiltonian oracle. This analysis would more complex than that for electronic structure due to the presence of bosons and fermions and the more complex form of the second quantized Hamiltonian. The analysis would also need to be repeated for each model considered, whereas the results we give here can be generalized more directly to any Hamiltonian in second quantized form.

We describe three quantum subroutines:
\begin{enumerate}
    \item A subroutine that enumerates the positions of the nonzero matrix elements of given row of the Hamiltonian in the Fock basis.  This subroutine is described in App.~\ref{enumerateH}, and requires $\widetilde{O}(\sqrt{K})$ local gates.
    \item A subroutine that, given a pair of Fock states each with total momentum $K$, computes the first $\bitprec$ bits of the matrix element connecting the states in an ancilla register. This subroutine is described in App.~\ref{computematrixelements}, and requires ${\widetilde{O}(\bitprec + K)}$ local gates.
    \item A subroutine that permits evaluation of the number operator for a given momentum mode (see also Sec.~\ref{measurement}). This subroutine is described in App.~\ref{occupationmeasurement}, and requires ${\widetilde{O}(\sqrt{K}+\bitprec)}$ local gates.
\end{enumerate}

\subsection{Matrix element enumeration\label{enumerateH}}

The Hamiltonian connects a pair of Fock states only if they are the same or the difference between them is exactly two fermions or antifermions (i.e., either two fermions, two antifermions, or a fermion and an antifermion) and either one or two bosons.
In other words, given a Fock state
\begin{equation}
    |\psi\rangle=|n_1,n_2,...,n_I\rangle\otimes|\overline{n}_1,\overline{n}_2,...,\overline{n}_I\rangle\otimes|(\widetilde{n}_1,\widetilde{\occ}_1),(\widetilde{n}_2,\widetilde{\occ}_2),...,(\widetilde{n}_I,\widetilde{\occ}_I)\rangle \ ,
\end{equation}
we may generate the states $|\psi\rangle$ is connected to by listing the possible changes the various terms in the Hamiltonian may make to $|\psi\rangle$.

We represent these possible changes as lists, which we will denote $\Delta$. The nonzero elements of a row or column in the Hamiltonian will be indexed by ordering the set of changes giving rise to the nonzero elements starting from the Fock state labeling the row. The $i$th nonzero element of the row will be labeled by $\Delta_i$. Each $\Delta$ has the form:
\begin{equation}
\label{change}
    \Delta=(k_1^+,t_1;~k_2^\pm,t_2;~k_3^\pm,t_3;~k_4^-,t_4) \ ,
\end{equation}
where $k_1^+$ is the momentum of the first momentum state whose occupancy will be increased by one (by a creation operator), and $t_1$ is indicates what type of particle it is (fermion, antifermion, or boson). Similarly, $k_2^\pm$ and $k_3^\pm$ are the momenta of the second and third momentum states whose occupancy is changed, which may be added or removed (since a term in the Hamiltonian possesses between one and three creation operators); $t_2$ and $t_3$ indicate their types. Finally, $k_4^-$ is the momentum of the fourth momentum state whose occupancy changes, which if present, must be lowered, since no term in the Hamiltonian contains four creation operators; $t_4$ indicates its type. Note that the ordering of increases and decreases in occupancy here is not that given by order of the creation and annihilation operators in the second quantized representation of Hamiltonian terms.

If one or more of these is not needed (because the change being described involves fewer than four particles), then the corresponding $k_j^\pm$ is set to zero. Thus either all four $k_j^\pm$ are nonzero (describing changes due to terms containing four ladder operators), only first three are nonzero (describing changes due to terms containing three ladder operators), or all $k_j^\pm$ are zero (describing the connection of $|\psi\rangle$ to itself via the number operators in \eqref{hmass}). Finally, in order to ensure that the $\Delta$ associated to any particular change is unique, we require that particles added appear first, followed by particles removed, and subject to that rule, types are ordered as fermions in increasing order of momentum, then antifermions in increasing order of momentum, then bosons in increasing order of momentum. This induces an ordering on the $\Delta$, which lets us enumerate the $\Delta_i$, and hence the nonzero matrix elements in a row.

Let the $t_i$ encode fermions, antifermions, and bosons as $0$, $1$, $2$, respectively.
Allowed changes in occupancy are then:
\begin{enumerate}
    \item $k_1^+=k_2^\pm=k_3^\pm=k_4^-=0$;
    \item $k_1^+,k_2^\pm,k_3^\pm\neq0$, $k_4^-=0$, and exactly one of $t_1,t_2,t_3$ is 2 (boson); or
    \item $k_1^+,k_2^\pm,k_3^\pm,k_4^-\neq0$, and exactly two of $t_1,t_2,t_3,t_4$ are 2 (bosons).
\end{enumerate}
In addition to these rules, the change must conserve momentum, i.e.,
\begin{equation}
\label{momentumconservation}
    k_1\pm k_2^{\pm}\pm'k_3^{\pm'}-k_4=0 \ .
\end{equation}

Since $k_i^\pm\in\{1,2,...,K\}$ (together with a bit encoding the sign) and $t_i\in\{0,1,2\}$ for each $i$, the number of distinct $\Delta$ appears to scale like $K^4$. However, the momentum conservation constraint \eqref{momentumconservation} means that one of the $k_i^\pm$ is determined by the other three, so in fact there are fewer than $3^4K^32^2=324K^3$  distinct $\Delta$s ($3^4$ possible valuations for the $t_i$, $K^3$ possible valuations for the $k_i$, and $2^2$ possible valuations for the two $\pm$s). This still overcounts the nonzero elements because the sparsity of the Hamiltonian in the Fock basis is $\Theta(K^2)$. This is because the full set of $\Delta$s considered here can act on certain states to produce unphysical occupancies - either fermion occupancies greater than one, boson occupancies above cutoff, or negative occupancies. Hence not every $\Delta$ returns a matrix element. We denote the number of $\Delta$ by $L=O(K^3)$, and index them as $\Delta_i$ for $0\leq i\leq L-1$.

\noindent
\textbf{Operations:}\\
We enumerate the $\Delta_i$ as follows. First we note that the Hamiltonian is a sum of $O(1)$ types of term labelled by tuples of momentum orbitals subject to momentum conservation constraints. For example, consider vertex terms of type $\OPbd_k\OPb_m\OPcd_l$, where $k=m+l$ and $3\leq k\leq K$. Given the tuple $k$, $m$ $l$ we can construct the $\Delta$ corresponding to this term with $O(\log K)$ operations. It only remains to show how to enumerate all tuples $(k,l,m)$. In fact, we only need to enumerate $(k,l)$ with $k>l$ and compute $m=k-l$. The first few tuples $(k,l)$ are $(2,1), (3,1), (3,2), (4,1)$. Let $n(k,l)$ be the number of the tuple $(k,l)$, with $n(2,1)=1$. Then $n(k,l)=(l-1)+n(k,1)$ and:
\begin{equation}
n(k,1)=1+\sum_{j=2}^{k-1} (j-1) = \frac{k(k-1)}{2}-(k-2)
\end{equation}
from which we obtain:
\begin{equation}
k(n) =\biggl\lfloor\frac{3+\sqrt{8n-7}}{2}\biggr\rfloor \ , \qquad l(n)=n-\frac{k(n)(k(n)-1)}{2}-(k(n)-2) \ .
\end{equation}
Hence $k,l,m$ can be computed from $n$ by $O(1)$ elementary arithmetic operations, the most costly of which is the square root. Therefore, the enumeration of the $\Delta_i$ corresponding to terms of type $\OPbd_k\OPb_m\OPcd_l$ requires $O((\log K)^2\log\log K)$ elementary gates. Similar arguments apply to seagull and fork terms.

We can now implement the oracle $O_F$ with action:
\begin{equation}
\label{fullmapping}
    O_F:|\psi\rangle\otimes|i\rangle\otimes|0\rangle_F~\mapsto~|\psi\rangle\otimes|i\rangle\otimes|\phi_i\rangle_F \ ,
\end{equation}
Where $i$ enumerates the fock states $\ket{\phi_i}$ such that $\bra{\psi}H\ket{\phi_i}\neq0$. The index $i$ runs over all types of terms in the Hamiltonian and all labellings of each type of term by momentum tuples as discussed above. The mapping \eqref{fullmapping} may be implemented in three steps, using an additional ancillary register capable of encoding a $\Delta$, also initialized to $0$:
\begin{equation}
\label{fullmappingsteps}
\begin{alignedat}{88}
    O_F:|\psi\rangle\otimes|i\rangle\otimes|0\rangle\otimes|0\rangle_F~
    &\mapsto~|\psi\rangle&&\otimes|i\rangle&&\otimes|\Delta_i\rangle&&\otimes|\psi\rangle_F\\
    &\mapsto~|\psi\rangle&&\otimes|i\rangle&&\otimes|\Delta_i\rangle&&\otimes|\phi_i\rangle_F\\
    &\mapsto~|\psi\rangle&&\otimes|i\rangle&&\otimes|0\rangle&&\otimes|\phi_i\rangle_F \ .
\end{alignedat}
\end{equation}
\begin{sloppypar}
The first step computes $\Delta_i$ from $i$ and copies the Fock state $|\psi\rangle$ to an ancilla register. Note that this Fock state $\psi$ is a computational basis state of the qubits and so the no-cloning theorem does not forbid this operation. The Fock state $|\phi_i\rangle$ is  obtained by changing $|\psi\rangle$ according to $\Delta_i$ if the resulting state $|\phi_i\rangle$ is physical, or $|\phi_i\rangle=|0\rangle$ if changing $|\psi\rangle$ according to $\Delta_i$ results in an unphysical state (for example, if $\Delta_i$ would remove a particle from a mode that is unoccupied in $|\psi\rangle$). Finally, we invert the first mapping.

We now describe the second step in this mapping in detail. There are two substeps. First, we must check whether $|\psi\rangle$ can be changed according to $\Delta_i$. We check that for each fermion and antifermion added by $\Delta_i$, the corresponding mode in $|\psi\rangle$ is empty, and for each particle removed by $\Delta_i$, the corresponding mode in $|\psi\rangle$ is nonempty. To determine this by a reversible computation that can be made coherent, we append to ${\Delta_i=(k_1^+,t_1;~k_2^\pm,t_2;~k_3^\pm,t_3;~k_4^-,t_4)}$ four ancillary qubits $|c_1,c_2,c_3,c_4\rangle$, initially all $0$, and for each particle change $(k_i^\pm,t_i)$, flip the corresponding bit $c_i$ if the particle change cannot be performed on $|\psi\rangle$. Thus if ${c_1=c_2=c_3=c_4=0}$, then $|\psi\rangle_F$ can be changed according to $\Delta_j$. If any one of the $c_i$ is nonzero, then $|\psi\rangle_F$ cannot be changed according to $\Delta_i$, so we set $\phi_i=0$. We first consider adding particles to $\ket{\psi}$, then consider removing them.

For each $(k_j^+,t_j)\in\Delta_i$, the mode is either present or absent in $\ket{\psi}$. This can be determined by $O(\sqrt{K})$ gates. If the mode is present and $t_j$ is $2$ (indicating that the added particle is a boson) then we simply increase its occupation by one. If the mode is not present, we must add it to $|\psi\rangle_F$. Because the modes must appear in order of increasing momentum, adding a new mode requires shifting all modes with particle type $t_j$ and momentum greater than $k_j$ over by $O(\log K)$ qubits. We append the new mode to the register containing particles of type $t_j$ above the highest momentum mode present, $k'$. We check if $k_j>k'$: if so, we have updated $\ket{\psi}$. If not, we exchange the new mode and mode $k'$, requiring $O(\sqrt{K})$ gates, and compare $k_j$ with the next smallest momentum mode. In the worst case the new mode has the smallest momentum and so this operation requires $O(\sqrt{K}\log K)$ gates.

For each $(k_j^-,t_j)\in\Delta_i$, we must remove a particle of type $t_j$ and momentum $k_j$. To do this, we reverse the method we used to add a mode, thus requiring $O(\sqrt{K}\log K)$ gates.   Beginning from the mode $k'$ of type $t_j$ with lowest momentum in $|\psi\rangle_F$, check whether $k_j=k'$: if it is, then if its initial occupation is greater than one, decrease its occupation by one.
If its initial occupation is one, remove this mode by setting the state of the corresponding mode register to 0, and then swap it to the end of the register. This completes the implementation of the second step in \eqref{fullmappingsteps}, which in turn completes the full mapping \eqref{fullmapping}.
\end{sloppypar}

\subsection{Computing matrix elements\label{computematrixelements}}

We take the first set of terms in $H_S$ \eqref{hseagull} as our example: we call this set of terms $H_{S,1}$.
The matrix elements due to the remaining sets of terms in the Hamiltonian may be computed using similar methods.
Substituting the explicit expressions for $c_\Moden$ and $\{\cdot,\cdot\}$ into the first line of \eqref{hseagull} gives:
\begin{equation}
	H_{S,1}=\sum_{\Modek,\Model,\Modem,\Moden}\frac{\delta_{\Modek+\Model,\Modem+\Moden}}{\sqrt{\Model\Moden}}\left(\frac{1-\delta_{\Modek,\Moden}}{\Modek-\Moden+\delta_{\Modek,\Moden}}+\frac{1}{\Modek+\Model}\right)(\OPbd_\Modek \OPb_\Modem \OPad_\Model \OPa_\Moden) \ .
\end{equation}
Note that the term $\delta_{\Modek,\Moden}$ appears in the denominator so that the first term is unambiguously zero if $k=n$. Assuming $|\psi\rangle$ and $|\psi'\rangle$ both have total momentum $K$, $\langle\psi'|H_{S,1}|\psi\rangle\neq0$ if and only if for
\begin{equation}
\label{inputstates}
\begin{alignedat}{8}
	|\psi\rangle&=|n_1,n_2,...,n_I\rangle\otimes|\overline{n}_1,\overline{n}_2,...,\overline{n}_I\rangle\otimes|(\widetilde{n}_1,\widetilde{\occ}_1),(\widetilde{n}_2,\widetilde{\occ}_2),...,(\widetilde{n}_I,\widetilde{\occ}_I)\rangle \ &&,\\
	|\psi'\rangle&=|n'_1,n'_2,...,n'_I\rangle\otimes|\overline{n}'_1,\overline{n}'_2,...,\overline{n}'_I\rangle\otimes|(\widetilde{n}'_1,\widetilde{\occ}'_1),(\widetilde{n}'_2,\widetilde{\occ}'_2),...,(\widetilde{n}'_I,\widetilde{\occ}'_I)\rangle \ &&,
\end{alignedat}
\end{equation}
the sets $\{n_1,n_2,...,n_I\}$ and $\{n'_1,n'_2,...,n'_I\}$ each contain exactly one element not in the other, the sets $\{\overline{n}_1,\overline{n}_2,...,\overline{n}_I\}$ and $\{\overline{n}'_1,\overline{n}'_2,...,\overline{n}'_I\}$ are identical, and
\begin{equation}
\begin{alignedat}{8}
	&\{(\widetilde{n}_1,\widetilde{\occ}_1),(\widetilde{n}_2,\widetilde{\occ}_2),...,(\widetilde{n}_I,\widetilde{\occ}_I)\}\ominus\{(\widetilde{n}'_1,\widetilde{\occ}'_1),(\widetilde{n}'_2,\widetilde{\occ}'_2),...,(\widetilde{n}'_I,\widetilde{\occ}'_I)\}\\
	&=\begin{cases}
		\{(\widetilde{n}_i,1),(\widetilde{n}'_j,1)\} \ &\text{s.t. } \ \widetilde{n}_i\neq\widetilde{n}'_j, \ \text{or}\\
		\{(\widetilde{n}_i,\widetilde{\occ}_i),(\widetilde{n}'_j,\widetilde{\occ}'_j),(\widetilde{n}'_{j'},1)\} \ &\text{s.t. } \  (\widetilde{n}'_{j\hphantom{'}},\widetilde{\occ}'_{j\hphantom{'}})=(\widetilde{n}_{i\hphantom{'}},\widetilde{\occ}_{i\hphantom{'}}-1), \ \text{or} \\
		\{(\widetilde{n}_i,\widetilde{\occ}_i),(\widetilde{n}'_j,\widetilde{\occ}'_j),(\widetilde{n}_{i'},1)\} \ &\text{s.t. } \ (\widetilde{n}'_{j\hphantom{'}},\widetilde{\occ}'_{j\hphantom{'}})=(\widetilde{n}_{i\hphantom{'}},\widetilde{\occ}_{i\hphantom{'}}+1), \ \text{or}\\
		\{(\widetilde{n}_i,\widetilde{\occ}_i),(\widetilde{n}'_j,\widetilde{\occ}'_j),(\widetilde{n}_{i'},\widetilde{\occ}_{i'}),(\widetilde{n}'_{j'},\widetilde{\occ}'_{j'})\}  &\text{s.t. } \
		\smash[b]
		{\begin{alignedat}[t]{8}
			& (\widetilde{n}'_{j\hphantom{'}},\widetilde{\occ}'_{j\hphantom{'}}) = (\widetilde{n}_{i\hphantom{'}},\widetilde{\occ}_{i\hphantom{'}}+1)
		    \\
		    \mathllap{\mathrlap{\text{and}}\hphantom{\text{s.t. } \ }}
		    & (\widetilde{n}'_{j'},\widetilde{\occ}'_{j'}) =(\widetilde{n}_{i'},\widetilde{\occ}_{i'}-1),
		\end{alignedat}}
	\end{cases}
\end{alignedat}\\[6pt]
\end{equation}
\vspace{0.25in}
for some set of indices $i,j,i',j'$ ($\ominus$ denotes symmetric difference).

\noindent
\textbf{Operations:}\\
Given the input states
\begin{equation}
	|\psi\rangle\otimes|\psi'\rangle \ ,
\end{equation}
we wish to evaluate the corresponding matrix element.
To do this, we attach ancillary registers whose state has the form
\begin{equation}
\label{ancillastate}
\begin{alignedat}{8}
	&\Big(|f_1,f_2,...,f_I\rangle\otimes|s\rangle\otimes|m\rangle\otimes|f'_1,f'_2,...,f'_I\rangle\otimes|s'\rangle\otimes|k\rangle\Big)\otimes\Big(|\overline{f}_1,\overline{f}_2,...,\overline{f}_I\rangle\otimes|\overline{s}\rangle\Big)\\
	&\otimes\Big(|(\widetilde{f}_1,\widetilde{g}_1),(\widetilde{f}_2,\widetilde{g}_2),...,(\widetilde{f}_I,\widetilde{g}_I)\rangle\otimes|\widetilde{s}\rangle\otimes|n\rangle\otimes|(\widetilde{f}'_1,\widetilde{g}'_1),(\widetilde{f}'_2,\widetilde{g}'_2),...,(\widetilde{f}'_I,\widetilde{g}'_I)\rangle\otimes|\widetilde{s}^{\,\prime}\rangle\otimes|l\rangle\otimes|\widetilde{\occ},\widetilde{\occ}'\rangle\Big) \ ,
\end{alignedat}
\end{equation}
where the $f_i,f'_j,\overline{f}_i,\widetilde{f}_i,\widetilde{f}'_j$ are each qubits initially set to $|1\rangle$. The $s,s',\overline{s},\widetilde{s},\widetilde{s}^{\,\prime}$ are each registers capable of encoding $I$ as a binary number. The $k,l,m,n,\widetilde{\occ},\widetilde{\occ}'$ are each registers capable of encoding $K$ as a binary number, initially set to 0. The $\widetilde{g}_i,\widetilde{g}'_j$ are each registers capable of encoding $K$ as a binary number, initially set to $1$.
We then perform the following operations:
\begin{enumerate}
    \item For each pair $(i,j)\in\{1,2,...,I\}^2$, perform the following mapping on the registers encoding $|n_i,n'_j,f_i,f'_j\rangle$ (initially in the state $|n_i,n'_j,1,1\rangle$):
    \begin{equation}
        |n_i,n'_j,1,1\rangle \ \mapsto \
        \begin{cases}
            |n_i,n'_j,0,0\rangle \ \text{if $n_i=n'_j$,}\\
            |n_i,n'_j,1,1\rangle \ \text{if $n_i\neq n'_j$.}
        \end{cases}
    \end{equation}
    There are $O(I^2)$ pairs and so the cost of this step is $O(K\log K)$ gates.
    \item For each $i=1,2,...,I$, perform the following mappings on the registers encoding $|f_i,s\rangle$ and $|f'_i,s'\rangle$:
    \begin{equation}
    \begin{alignedat}{8}
        &\sket{f_i,s} \ &&\mapsto \ && \sket{f_i,s+f_i}&& \ ,\\
        &\sket{f'_i,s'} \ &&\mapsto \ && \sket{f'_i,s'+f'_i}&& \ .
    \end{alignedat}
    \end{equation}
    The cost of this step is $O(\sqrt{K}\log K)$ gates.

    \item For each $i=1,2,...,I$, perform the following mappings on the registers encoding $|f_i,n_i,m\rangle$ and $|f'_i,n'_i,k\rangle$:
    \begin{equation}
    \begin{alignedat}{8}
        &|f_i,n_i,m\rangle \ &&\mapsto \ &&|f_i,n_i,m+f_in_i\rangle&& \ ,\\
        &|f'_i,n'_i,k\rangle \ &&\mapsto \ &&|f'_i,n'_i,k+f'_in'_i\rangle&& \ .
    \end{alignedat}
    \end{equation}
    The cost of this step is $O(\sqrt{K}\log K)$ gates.

    \item For each pair $(i,j)\in\{1,2,...,I\}^2$, perform the following mapping on the registers encoding $|\overline{n}_i,\overline{n}'_j,\overline{f}_i\rangle$ (initially in the state $|\overline{n}_i,\overline{n}'_j,1\rangle$):
    \begin{equation}
        |\overline{n}_i,\overline{n}'_j,1\rangle \ \mapsto \
        \begin{cases}
            |\overline{n}_i,\overline{n}'_j,0\rangle \ \text{if $\overline{n}_i=\overline{n}'_j$,}\\
            |\overline{n}_i,\overline{n}'_j,1\rangle \ \text{if $\overline{n}_i\neq\overline{n}'_j$.}
        \end{cases}
    \end{equation}
    The cost of this step is $O(K\log K)$ gates.

    \item For each $i=1,2,...,I$, perform the following mapping on the registers encoding $|\overline{f}_i,\overline{s}\rangle$:
    \begin{equation}
        |\overline{f}_i,\overline{s}\rangle \ \mapsto \ |\overline{f}_i,\overline{s}+\overline{f}_i\rangle \ .
    \end{equation}
    The cost of this step is $O(\sqrt{K}\log K)$ gates.

    \item For each pair $(i,j)\in\{1,2,...,I\}^2$, perform the following mapping on the registers encoding $|(\widetilde{n}_i,\widetilde{\occ}_i),(\widetilde{n}'_j,\widetilde{\occ}'_j),\widetilde{f}_i,\widetilde{g}_i\rangle$ (initially in the state $|(\widetilde{n}_i,\widetilde{\occ}_i),(\widetilde{n}'_j,\widetilde{\occ}'_j),1,1\rangle$):
    \begin{equation}
    \label{bosonprelimmapping1}
    \begin{alignedat}{8}
        &|(\widetilde{n}_i,\widetilde{\occ}_i),(\widetilde{n}'_j,\widetilde{\occ}'_j),1,1\rangle\\
        & \ \mapsto \
        \left\{
        \begin{alignedat}{9}
            &|(\widetilde{n}_i,\widetilde{\occ}_i),(\widetilde{n}'_j,\widetilde{\occ}'_j),1,1\rangle \ &&\text{if $\widetilde{n}_i\neq\widetilde{n}'_j$,}\\
            &|(\widetilde{n}_i,\widetilde{\occ}_i),(\widetilde{n}'_j,\widetilde{\occ}'_j),0,0\rangle \ &&\text{if $\widetilde{n}_i=\widetilde{n}'_j$, and $\widetilde{\occ}_i=\widetilde{\occ}'_j\text{ or }\widetilde{\occ}'_j-1$,}\\
            &|(\widetilde{n}_i,\widetilde{\occ}_i),(\widetilde{n}'_j,\widetilde{\occ}'_j),1,\widetilde{\occ}_i\rangle \ &&\text{if $\widetilde{n}_i=\widetilde{n}'_j$, and $\widetilde{\occ}_i=\widetilde{\occ}'_j+1$,}\\
            &|(\widetilde{n}_i,\widetilde{\occ}_i),(\widetilde{n}'_j,\widetilde{\occ}'_j),1,0\rangle \ &&\text{if $\widetilde{n}_i=\widetilde{n}'_j$, and $|\widetilde{\occ}_i-\widetilde{\occ}'_j|>1$.}
        \end{alignedat}\right.
    \end{alignedat}
    \end{equation}
    The cost of this step is $O(K\log K)$ gates.

    \item For each pair $(i,j)\in\{1,2,...,I\}^2$, perform the following mapping on the registers encoding $|(\widetilde{n}_i,\widetilde{\occ}_i),(\widetilde{n}'_j,\widetilde{\occ}'_j),\widetilde{f}'_j,\widetilde{g}'_j\rangle$ (initially in the state $|(\widetilde{n}_i,\widetilde{\occ}_i),(\widetilde{n}'_j,\widetilde{\occ}'_j),1,1\rangle$):
    \begin{equation}
    \label{bosonprelimmapping2}
    \begin{alignedat}{8}
        &|(\widetilde{n}_i,\widetilde{\occ}_i),(\widetilde{n}'_j,\widetilde{\occ}'_j),1,1\rangle\\
        & \ \mapsto \
        \left\{
        \begin{alignedat}{9}
            &|(\widetilde{n}_i,\widetilde{\occ}_i),(\widetilde{n}'_j,\widetilde{\occ}'_j),1,1\rangle \ &&\text{if $\widetilde{n}_i\neq\widetilde{n}'_j$,}\\
            &|(\widetilde{n}_i,\widetilde{\occ}_i),(\widetilde{n}'_j,\widetilde{\occ}'_j),0,0\rangle \ &&\text{if $\widetilde{n}_i=\widetilde{n}'_j$, and $\widetilde{\occ}_i=\widetilde{\occ}'_j\text{ or }\widetilde{\occ}'_j+1$,}\\
            &|(\widetilde{n}_i,\widetilde{\occ}_i),(\widetilde{n}'_j,\widetilde{\occ}'_j),1,\widetilde{\occ}'_j\rangle \ &&\text{if $\widetilde{n}_i=\widetilde{n}'_j$, and $\widetilde{\occ}_i=\widetilde{\occ}'_j-1$,}\\
            &|(\widetilde{n}_i,\widetilde{\occ}_i),(\widetilde{n}'_j,\widetilde{\occ}'_j),1,0\rangle \ &&\text{if $\widetilde{n}_i=\widetilde{n}'_j$, and $|\widetilde{\occ}_i-\widetilde{\occ}'_j|>1$.}
        \end{alignedat}\right.
    \end{alignedat}
    \end{equation}
    The cost of this step is $O(K\log K)$ gates.

    \item For each $i=1,2,...,I$, perform the following mappings on the registers encoding $|\widetilde{f}_i,\widetilde{g}_i,\overline{s}\rangle$ and $|\widetilde{f}'_i,\widetilde{g}'_i,\overline{s}\rangle$:
    \begin{equation}
    \begin{alignedat}{8}
        &|\widetilde{f}_i,\widetilde{g}_i,\overline{s}\rangle \ &&\mapsto \ &&|\widetilde{f}_i,\widetilde{g}_i,\overline{s}+\widetilde{f}_i\delta_{\widetilde{g}_i,0}\rangle \ &&,\\
        &|\widetilde{f}'_i,\widetilde{g}'_i,\overline{s}\rangle \ &&\mapsto \ &&|\widetilde{f}'_i,\widetilde{g}'_i,\overline{s}+\widetilde{f}'_i\delta_{\widetilde{g}'_i,0}\rangle \ &&.
    \end{alignedat}
    \end{equation}
    The cost of this step is $O(\sqrt{K}\log K)$ gates.

    \item For each $i=1,2,...,I$, perform the following mappings on the registers encoding $|\widetilde{f}_i,\widetilde{s}\rangle$ and $|\widetilde{f}'_i,\widetilde{s}^{\,\prime}\rangle$:
    \begin{equation}
    \begin{alignedat}{8}
        &|\widetilde{f}_i,\widetilde{s}\rangle \ &&\mapsto \ &&|\widetilde{f}_i,\widetilde{s}+\widetilde{f}_i\rangle \ &&,\\
        &|\widetilde{f}'_i,\widetilde{s}^{\,\prime}\rangle \ &&\mapsto \ &&|\widetilde{f}'_i,\widetilde{s}^{\,\prime}+\widetilde{f}'_i\rangle \ &&,\\.
    \end{alignedat}
    \end{equation}
    The cost of this step is $O(\sqrt{K}\log K)$ gates.

    \item For each $i=1,2,...,I$, perform the following mappings on the registers encoding $|\widetilde{f}_i,\widetilde{n}_i,l\rangle$ and $|\widetilde{f}'_i,\widetilde{n}'_i,n\rangle$:
    \begin{equation}
    \begin{alignedat}{8}
        &|\widetilde{f}_i,\widetilde{n}_i,l\rangle \ &&\mapsto \ &&|\widetilde{f}_i,\widetilde{n}_i,l+\widetilde{f}_i\widetilde{n}_i\rangle \ &&,\\
        &|\widetilde{f}'_i,\widetilde{n}'_i,n\rangle \ &&\mapsto \ &&|\widetilde{f}'_i,\widetilde{n}'_i,n+\widetilde{f}'_i\widetilde{n}'_i\rangle \ &&,\\.
    \end{alignedat}
    \end{equation}
    The cost of this step is $O(\sqrt{K}\log K)$ gates.

    \item For each $i=1,2,...,I$, perform the following mappings on the registers encoding $|\widetilde{f}_i,\widetilde{\occ}_i,\widetilde{\occ}\rangle$ and $|\widetilde{f}'_i,\widetilde{\occ}'_i,\widetilde{\occ}'\rangle$:
    \begin{equation}
    \begin{alignedat}{8}
        &|\widetilde{f}_i,\widetilde{\occ}_i,\widetilde{\occ}\rangle \ &&\mapsto \ &&|\widetilde{f}_i,\widetilde{\occ}_i,\widetilde{\occ}+\widetilde{f}_i\widetilde{\occ}_i\rangle \ &&,\\
        &|\widetilde{f}'_i,\widetilde{\occ}'_i,\widetilde{\occ}'\rangle \ &&\mapsto \ &&|\widetilde{f}'_i,\widetilde{\occ}'_i,\widetilde{\occ}'+\widetilde{f}'_i\widetilde{\occ}'_i\rangle \ &&,\\.
    \end{alignedat}
    \end{equation}
    The cost of this step is $O(\sqrt{K}\log K)$ gates.
\end{enumerate}

\begin{sloppypar}
When steps 2 and 3 have been completed, the fermionic part of the condition for the states to be connected is satisfied if and only if ${s=s'=1}$, i.e., if exactly one of the $f_i$ and one of the $f'_j$ is 1.
$k$ and $m$ store the fermionic momenta whose occupations are changed in the case when the states may be connected.

When steps 4 and 5 have been completed, the antifermionic part of the condition for the states to be connected is satisfied if and only if $\overline{s}=0$.
When steps 6 and 7 have been completed, if the final case in \eqref{bosonprelimmapping1} or \eqref{bosonprelimmapping2} holds for any $i$ (or $j$), then the two states are not connected.
Step 8 therefore adds at least 1 to $\overline{s}$ if and only if the final case in \eqref{bosonprelimmapping1} or \eqref{bosonprelimmapping2} holds for some $i$ or $j$.
Thus after these operations are complete, the states can only be connected if $\overline{s}=0$.

When step 9 has been completed, the states can be connected only when ${\widetilde{s}=\widetilde{s}^{\,\prime}=1}$, i.e., if exactly one of the $\widetilde{f}_i$ and one of the $\widetilde{f}'_j$ is 1.
Thus when step 10 has been completed, $l$ and $n$ store the bosonic momenta whose occupations are changed in the case when the states may be connected; and when step 11 has been completed, $\widetilde{\occ},\widetilde{\occ}'$ store the (larger) occupation numbers of the two bosonic momenta that change between the two states.
\end{sloppypar}

Having implemented all of the preceding operations, the matrix element $\langle\psi'|H_{S,1}|\psi\rangle$ may be computed as follows:
\begin{enumerate}
	\item $\langle\psi'|H_{S,1}|\psi\rangle\neq0$ if and only if $s=s'=\widetilde{s}=\widetilde{s}^{\,\prime}=1$ and $\overline{s}=0$.
	
	\item If the above condition holds, then
		\begin{equation}
			\label{matrixelement}
			\begin{split}
			    \langle\psi'|H_{S,1}|\psi\rangle&=\sqrt{\frac{\widetilde{\occ}\widetilde{\occ}'}{ln}}\left(\frac{1-\delta_{k,n}}{k-n+\delta_{k,n}}+\frac{1}{k+l}\right)\\
			    &=
			    \sqrt{\frac{\widetilde{\occ}\widetilde{\occ}'}{ln}}\times
			    \begin{cases}
                \dfrac{1}{k-n}+\dfrac{1}{k+l} \ &\text{if $k\neq n$,}  \\
                    \dfrac{1}{k+l} \ &\text{if $k=n$.}
			    \end{cases}
			\end{split}
		\end{equation}
		
	\item The matrix element \eqref{matrixelement} is a function of the numbers $k,l,m,n,\widetilde{\occ},\widetilde{\occ}'$, each of which has already been stored in its own register of $\lceil\log_2K\rceil$ qubits. Thus we may compute the matrix element to any desired number of bits $\bitprec$ and store it in a register of the same length, by an operation on the $6\lceil\log_2K\rceil+\bitprec$ qubits involved.
	Computation of the square root requires two multiplications of two $O(\log{K})$-bit numbers, costing $O(\log{K}^2)$ gates. These two numbers are then divided, yielding a result with $O(\bitprec)$ bits of precision, requiring $O(\bitprec^2)$ operations. We then take the square root of this $\bitprec$-bit number, requiring $O(\bitprec^2\log \bitprec)$ gates. To compute the second term in the case $k\neq n$ we can either perform one addition and one subtraction of two $O(\log K)$-bit numbers, followed by two divisions, or compute the common denominator and numerator, and perform one division. In either case the cost is $O(\log K + \bitprec^2)$. Thus calculating the matrix element requires
	\begin{equation}
	\label{matxcost}
	    O\bigl((\log K)^2 + \bitprec^2\log \bitprec\bigr)
	\end{equation}
	gates.
\end{enumerate}

The matrix elements due to other terms in the Hamiltonian may be evaluated using similar methods.
Similar analyses will apply for each term, so the overall cost to evaluate a matrix element of the full Hamiltonian will be
\begin{equation}
	O(K\log K + \bitprec^2\log \bitprec)=\widetilde{O}(\bitprec^2 + K)
\end{equation}
gates, where the dependence on $\bitprec$ comes from the final calculation of the matrix element.

\begin{sloppypar}
We can also calculate the total number of qubits required for these operations.
The input states in \eqref{inputstates} each require $I\lceil\log_2K\rceil$ qubits for fermions, $I\lceil\log_2K\rceil$ qubits for antifermions, and $2I\lceil\log_2K\rceil$ qubits for bosons, for a total of
\begin{equation}
\label{qubitsforinput}
    8I\lceil\log_2K\rceil\le8\sqrt{2K}\lceil\log_2K\rceil\le12\sqrt{K}\lceil\log_2K\rceil
\end{equation}
\begin{sloppypar*}
qubits to encode the input states; as we would hope (since we have two input states) this is twice the number required to encode a single state in the compact mapping, as described in App.~\ref{partitions}, eq.~\eqref{singlestatetotalqubits}.
The ancillary registers in \eqref{ancillastate} require $5I$ qubits for the ${\{|f_i\rangle,|f'_j\rangle,|\overline{f}_i\rangle,|\widetilde{f}_i\rangle,|\widetilde{f}'_j\rangle\}}$, ${5\lceil\log_2I\rceil}$ qubits for ${\{|s\rangle,|s'\rangle,|\overline{s}\rangle,|\widetilde{s}\rangle,|\widetilde{s}^{\,\prime}\rangle\}}$, ${6\lceil\log_2K\rceil}$ qubits for ${\{|k\rangle,|l\rangle,|m\rangle,|n\rangle,|\widetilde{\occ}\rangle,|\widetilde{\occ}'\rangle\}}$, and ${2I\lceil\log_2K\rceil}$ qubits for the ${\{\widetilde{g}_i,\widetilde{g}'_j\}}$, for a total of
\end{sloppypar*}
\begin{equation}
\label{ancillaqubits}
    5I+5\lceil\log_2I\rceil+6\lceil\log_2K\rceil+2I\lceil\log_2K\rceil\le2\sqrt{2K}\lceil\log_2K\rceil+5\sqrt{2K}+11\lceil\log_2K\rceil+\bitprec
\end{equation}
qubits.
Thus the total number of qubits required is upper bounded by
\begin{equation}
\label{totalqubits}
    10\sqrt{2K}\lceil\log_2K\rceil+5\sqrt{2K}+11\lceil\log_2K\rceil\le15\sqrt{K}\lceil\log_2K\rceil+8\sqrt{K}+11\lceil\log_2K\rceil+\bitprec
\end{equation}
(where, again, $\bitprec$ is the number of bits desired in the output matrix element).
\end{sloppypar}

\subsection{Measurement of an occupation number\label{occupationmeasurement}}

In order to evaluate PDFs, we need to be able to measure the number operator $\numberop_\ell$, to estimate the expectation value \eqref{pdfour}. This means that for an encoded Fock state
\begin{equation}
    |n_1,n_2,...,n_I\rangle\otimes|\overline{n}_1,\overline{n}_2,...,\overline{n}_I\rangle\otimes|(\widetilde{n}_1,\widetilde{\occ}_1),(\widetilde{n}_2,\widetilde{\occ}_2),...,(\widetilde{n}_I,\widetilde{\occ}_I)\rangle \ ,
\end{equation}
we want to perform a measurement of the occupation number of some particular momentum mode $\Moden$, summed over fermions, antifermions, and bosons.

To do this, we employ an ancillary register whose states
have the form:
\begin{equation}
    |f_1,f_2,...,f_I\rangle\otimes|\overline{f}_1,\overline{f}_2,...,\overline{f}_I\rangle\otimes|\widetilde{f}_1,\widetilde{f}_2,...,\widetilde{f}_I\rangle\otimes|s\rangle \ ,
\end{equation}
where the $f_i,\overline{f}_i,\widetilde{f}_i$ are each qubits initially set to $|0\rangle$, and $s$ is a register of $O(\log K)$ qubits, initially set to $0$. For the desired $\Moden$, we iterate over the $n_i$, $\overline{n}_i$, and $\widetilde{n}_i$, checking whether each is equal to $\Moden$: if it is, then we set the corresponding $f_i$, $\overline{f}_i$, or $\widetilde{f}_i$ to 1.
This requires $3I$ operations on $O(\log K)$ qubits, requiring $O(I\log K)$ gates.

Now, we iterate over the $f_i$ and $\overline{f}_i$, adding their values to $s$. Each such operation is a binary addition on $O(\log K)$ qubits, and we implement $2I$ of them, so the total number of gates required is again $O(I\log K)$.
After this step is complete, $s$ will encode the total number of fermions and antifermions with momentum $\Moden$ (between $0$ and $2$).

Finally, we iterate over the pairs $(\widetilde{\occ}_i,\widetilde{f}_i)$, adding the products of their values to $s$, i.e.,
\begin{equation}
    \label{countnumber}
    s \ \mapsto \ s+\sum_{i=1}^I\widetilde{\occ}_i\widetilde{f}_i \ .
\end{equation}
Each such operation is a binary addition on $O(\log K)$ qubits, and we implement $I$ of them, so the total number of gates required is again $O(I\log K)$. After this step is complete, $s$ will encode the total number of fermions, antifermions, and bosons with momentum $\Moden$.
Once this routine is complete, we can sample the occupation number of mode $\Moden$ by measuring the qubits encoding $s$.
The total cost is $O(I\log K)=\widetilde{O}(\sqrt{K})$.

The situation becomes only slightly more complicated when we impose the probing scale~$\Qtransf^2$ as in~\eqref{pokeQ}.
Now we wish to estimate the expectation value of the number operator $\numberop_\ell$ for the cutoff state $|\Psi^{(\Qtransf)}\rangle$, as in~\eqref{pdfourcutoff}.
We use the following version of the cutoff condition \eqref{pokeQ}:
\begin{equation}
\begin{alignedat}{8}
    \sum_{j=1}^I\left(\frac{m_j^2}{n_j}+\frac{\overline{m}_j^2}{\overline{n}_j}+\widetilde{\occ}_j\frac{\widetilde{m}_j^2}{\widetilde{n}_j}\right)\le\frac{\Qtransf^2}{K} \ .
\end{alignedat}
\end{equation}
To calculate the left-hand of this expression, we employ an additional ancillary register whose states have the form:
\begin{equation}
    |s'\rangle\otimes|t\rangle \ ,
\end{equation}
where $|s'\rangle$ is a register of $\bitprec$ qubits (which we will use to store a floating point number, initially~$0$), and $|t\rangle$ is a single qubit.

To evaluate the cutoff condition:
\begin{enumerate}
    \item For each $j=1,2,...,I$, perform the following mappings on the registers encoding $|n_j,s'\rangle$ and $|\overline{n}_j,s'\rangle$:
    \begin{equation}
    \begin{alignedat}{8}
        |n_j,s'\rangle \ &&\mapsto \ &&|n_j,s'+\frac{m_j^2}{n_j}\rangle \ ,\\
        |\overline{n}_j,s'\rangle \ &&\mapsto \ &&|\overline{n}_j,s'+\frac{\overline{m}_j^2}{\overline{n}_j}\rangle \ .
    \end{alignedat}
    \end{equation}

    \item For each $j=1,2,...,I$, perform the following mappings on the registers encoding $|(\widetilde{n}_j,\widetilde{\occ}_j),s'\rangle$:
    \begin{equation}
        |(\widetilde{n}_j,\widetilde{\occ}_j),s'\rangle \ \mapsto \ |(\widetilde{n}_j,\widetilde{\occ}_j),s'+\widetilde{\occ}_j\frac{\widetilde{m}_j^2}{\widetilde{n}_j}\rangle \ .
    \end{equation}

    \item Perform the following mapping on the registers encoding $|s',t\rangle$ (which will be in the state $|s',0\rangle$ for some $s'$ set by the previous steps):
    \begin{equation}
        |s',0\rangle \ \mapsto \
        \begin{cases}
            |s',0\rangle \ \text{if }s'>{\Qtransf^2}/{K},\\
            |s',1\rangle \ \text{if }s'\le{\Qtransf^2}/{K}.
        \end{cases}
    \end{equation}
\end{enumerate}

When steps 1 and 2 have been completed, $s'$ will be an encoding (to the desired precision, set by its number of qubits $\bitprec$) of
\begin{equation}
    s'\approx\sum_{j=1}^I\left(\frac{m_j^2}{n_j}+\frac{\overline{m}_j^2}{\overline{n}_j}+\widetilde{\occ}_j\frac{\widetilde{m}_j^2}{\widetilde{n}_j}\right) \ .
\end{equation}
Thus the third step merely checks whether the value of $s'$ is bounded or not by the quantity~$\frac{\Qtransf^2}{K}$, which is classically precomputed, and updates the qubit $t$ accordingly.
Then in order to get the expectation value of the number operator for the cutoff state $|\Psi^{(\Qtransf)}\rangle$, as in \eqref{pdfourcutoff}, we compute the number operator as above for the non-cutoff state, but only for pairs $(s,t)$ where $t=1$; when $t=0$ we throw away the sample.

This avoids sampling values corresponding to disallowed states.
Note that if in some superposition of Fock states, those above the cutoff $\frac{\Qtransf^2}{K}$ possess too much of the total probability, it may become inefficient to sample only from the allowed states.
This situation can be avoided by keeping the imposed cutoff~$\Qtransf^2$ not too far below the maximum energy scale~$\Qtransf_{\max}^2(K)$; see Fig.~\ref{fig:PDFevol} and the associated discussion.

\begin{sloppypar}
Each of the operations in step 1 above involves a division and an addition on $\bitprec+\log_2 K$ qubits: $\log_2 K$ for the $n_j$ or $\overline{n}_j$, and $\bitprec$ for $s'$.
Each of the operations in step 2 above involves a division, a multiplication, and an addition on $\bitprec+2\log_2 K$ qubits: $\log_2 K$ for each of $\widetilde{n}_j$ and $\widetilde{\occ}_j$, and $\bitprec$ for $s'$.
We perform $2I$ of the first type, and $I$ of the second;
the final step is just a multiply-controlled NOT on $\bitprec+1$ qubits, so the total number of CNOTs and single-qubit gates required is $\widetilde{O}(I+\bitprec)=\widetilde{O}(\sqrt{K}+\bitprec)$.
\end{sloppypar}


\end{appendices}

\bibliography{main}
\bibliographystyle{unsrt}

\end{document}